# Low driving-force stable bending cooling via fatigue-resistant hierarchical NiTi shape memory alloy

Kai Yan[1,2], Kangjie Chu[1,2], Peng Hua[3], Pengbo Wei[1,2], Hanlin Gu[4], Qiming Zhuang[1], Weifeng He[5,6], Fuzeng Ren[1,*], Qingping Sun[2,*], Robert O. Ritchie[7,8,*]

[1]Department of Materials Science and Engineering, Southern University of Science and Technology, Shenzhen, China
[2]Department of Mechanical and Aerospace Engineering, The Hong Kong University of Science and Technology, Hong Kong, China
[3]School of Science, Harbin Institute of Technology, Shenzhen, China
[4]Department of Mechanics and Engineering Science, Peking University, Beijing, China
[5]State Key Laboratory for Manufacturing Systems Engineering, School of Mechanical Engineering, Xi'an Jiaotong University, Xi'an, China
[6]Science and Technology on Plasma Dynamics Laboratory, Air Force Engineering University, Xi'an, China
[7]Department of Materials Science and Engineering, University of California, Berkeley, CA, USA
[8]Materials Sciences Division, Lawrence Berkeley National Laboratory, Berkeley, CA, USA
*Corresponding authors. Email: renfz@sustech.edu.cn (F.R.); meqpsun@ust.hk (Q.S.); roritchie@lbl.gov (R.O.R)

**Abstract:**

Elastocaloric cooling with shape memory alloys (SMAs) is emerging as a promising candidate for next-generation, environmentally friendly refrigeration. However, its development is hindered by the large driving force and low efficiency associated with uniaxial loading modes. In response, we present an innovative elastocaloric air cooling approach that utilizes the bending of NiTi beams, offering a low-force and energy-efficient solution. We achieve a continuous maximum temperature drop of 0.91 K and a dissipated energy ($\Delta W$) of 15.1 N·mm at a low specific driving force of 220 N. Notably, the specimen achieves over 5 million cycles under a maximum surface tensile strain of 1.94% for macroscopic cyclic bending, via a pre-strained, warm laser shock peening (pw-LSP) method. This unprecedented fatigue resistance originates from the formation of a hierarchical microstructure and a large compressive residual stress of over 1 GPa. This work demonstrates the great potential of bending induced elastocaloric cooling in the near future.

**One-Sentence Summary:**

Yan et al. have developed a strategy to substantially enhance the bending fatigue life and elastocaloric performance of NiTi shape memory alloys, involving a continuous maximum temperature drop of 0.91 K with a dissipated energy ($\Delta W$) of 15.1 N·mm at a low specific driving force of 220 N, to achieve a life of over 5 million cycles under macroscopic bending with a maximum surface tensile strain of 1.94%, demonstrating the potential of this fatigue-resistant, hierarchical structure for low-force, energy-efficient elastocaloric cooling applications.

The polycrystalline NiTi shape memory alloy (SMA) is widely used in various fields such as biomedical devices, actuators, and elastocaloric refrigerators due to its superelasticity, strong corrosion resistance, large reversible deformation, and high damping capacity (*1*). It is suitable for energy conversion in solid-state cooling due to its large latent heat created by the stress-induced first-order phase transition (*2-6*). However, the fatigue performance of commercial components made of NiTi SMA is unsatisfactory due to crack nucleation occurring far below the macroscopic plastic flow stress of the material during cyclic loading under tensile and bending loading conditions (*7-9*). Failure of NiTi SMA typically involves both a reversible B2↔B19′ phase transition and dislocation slip, which contribute to cracking at the surfaces and grain/phase boundaries of the material under external cyclic stresses (*10*). Numerous factors, including grain size (*8*), precipitates (*11*), geometric compatibility (*12*), impurity levels (*13*), surface condition (*14*), testing environment (*15*), and loading conditions (*16, 17*), can affect the fatigue behavior of NiTi SMA. Surface compressive residual stress can enhance structural fatigue resistance, while grain refinement and introducing precipitates like $Ni_4Ti_3$ can enhance functional fatigue resistance (*18-23*). However, the fracture toughness and resistance to crack propagation of NiTi notably drop with grain size reduction (*24*). As a result, the tensile and bending fatigue lives of bulk nanocrystalline NiTi remain limited.

Surface nano-crystallization effectively circumvents the trade-off between crack nucleation resistance and crack propagation resistance in nanocrystalline NiTi, leading to a considerable increase in bending fatigue life (*25-27*). Previous studies by Liao *et al*. (*28, 29*) employed laser shock peening (LSP) and post-deformation annealing to induce residual martensite, resulting in a bimodal grain size distribution at the surface of a coarse-grained NiTi alloy. Wang *et al*. (*30*) introduced a dislocation structure and amorphization in a coarse-grained NiTi also using the LSP technique. Nevertheless, there has been only limited focus on the application of the LSP treatment for nanocrystalline NiTi; indeed, the intricate interactions between the laser and nanocrystalline NiTi under ultrahigh strain and temperature rates remain poorly understood.

We here develop a strategy to enhance the bending fatigue resistance of nanocrystalline NiTi SMA through the formation of a hierarchical microstructure and the introduction of a large compressive residual stress using pre-strain warm laser shock peening (pw-LSP) (fig. S1). The commercial NiTi shape memory alloy, initially in a B2 phase, underwent pre-straining to 10% and was then heated to 150 °C to achieve complete martensitic phase prior to undergoing Laser

Shock Peening (LSP) processing. This approach effectively bypassed the phase transition, enabling direct induction of plastic deformation in the martensite (fig. S4). During pw-LSP processing, the specimen surface experienced heat flux and pressure. This led to the formation of a hierarchical structure near the SMA surface, comprising a nanocomposite layer, a nitride layer, and an ultrafine-grained layer (Fig. 1A); in fact, transmission electron microscopy (TEM) observations validate the formation of such a structure. The topmost 20 nm featured a thin crystalline-amorphous nanocomposite layer (Fig. 1, B to D) with nanograined TiO, $Ni_4Ti_3$, and the B19′ phases distributed within the amorphous matrix (fig. S8). Beneath this layer lay the nitride layer, characterized by $Ti_2Ni$-rich dendrites in the top region (20 nm to 200 nm) (fig. S9, and table S1), snowflake-shaped coherent TiN-rich precipitates in the middle region (200 nm to 1 μm) (figs. S10 and S11, and table S2), and spherical $Ti_2N$-rich precipitates in the bottom region (1 μm to 2.5 μm) of the $Ni_{2.67}Ti_{1.33}$ matrix (figs. S12 and S13).

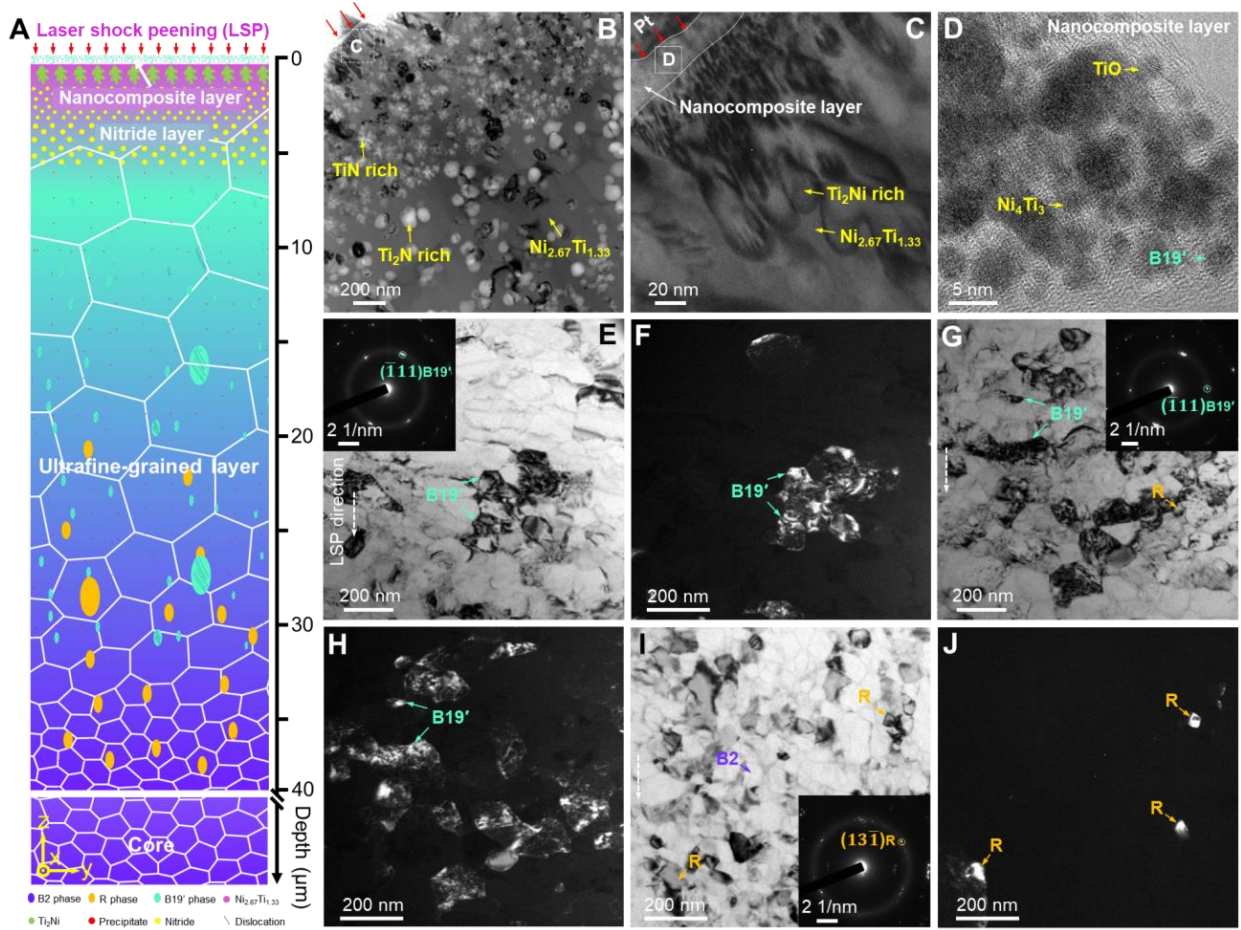

**Fig. 1. Hierarchical microstructure of NiTi SMA.** (**A**) Schematic illustration of the hierarchical microstructure of the NiTi. (**B**) TEM image of the LSP-treated surface. (**C**) Enlarged

TEM image of the LSP-treated surface. (**D**) HRTEM image of the nanocomposite layer. (**E-F**) Bright-field TEM image, SAED pattern, and the dark-field TEM image taken at a depth of 13 μm from the pw-LSP-treated surface. (**G-H**) Bright-field TEM image, SAED pattern, and dark-field TEM image taken at a depth of 20 μm from the pw-LSP-treated surface. (**I-J**) Bright-field TEM image, SAED pattern, and dark-field TEM image taken at a depth of 40 μm from the pw-LSP-treated surface.

Below the nitride layer, an ultrafine-grained region with phase evolution was observed. The grain growth in this region is evident, with some grains even reaching up to 900 nm (fig. S14A). Between 13 μm and 20 μm, a large number of B19′ phase nanodomains with $Ni_4Ti_3$ precipitates, and high-density dislocations were observed (Fig. 1, E to F, and fig. S14). From 20 μm to 40 μm, a two-phase coexistence region of R and B19′ phase was present, with numerous nanograins featuring a significant presence of dislocations (Fig. 1G). Selected area electron diffraction (SAED) confirmed the additional presence of $Ni_4Ti_3$ precipitates, while dark-field TEM imaging (Fig. 1H) revealed widespread B19′ phase nanodomains. At a depth of 40 μm (Fig. 1I), the microstructure reverted to untreated equiaxed grains, showing minimal plastic deformation and few dislocations. The grain size matched that of the as-received sample (~65 nm), but more phases were present. SAED patterns confirmed the presence of B2, R phase, and $Ni_4Ti_3$, with nanoscale R phase visualized in dark-field TEM imaging (Fig. 1J). Additionally, scanning probe microscope (SPM) imaging mapped the phase and modulus of different depths from the pw-LSP treated top surface, revealing the hierarchical microstructure (figs. S15 and S16).

Four-point bending tests were performed to study the bending fatigue behavior of the NiTi SMA. Calculated tensile and compressive strains at the specimen surface were approximately ± 1.94%. The evolution of the force-displacement curves during various loading cycles for the as-received nanocrystalline NiTi SMA showed unstable stress-strain responses and a short bending fatigue life, terminating in fracture after 1654 cycles (Fig. 2A and movie S1). In contrast, the hierarchical NiTi SMA demonstrated an exceptionally long fatigue life, approximately three thousand times that of the as-received counterpart. The sample remained intact even after 5 million cycles of cyclic bending (Fig. 2B and movie S2), although some functional degradation occurred due to the accumulation of dislocations and the formation of residual martensite and the R phase. During cyclic deformation, the buildup of dislocations cannot be recovered, resulting in the eventual fracture of the material. In addition, the remaining martensite and R phases can be partially recovered by heating at 100 °C. The infrared images of the nanocrystalline (movie S3) and hierarchical specimens (movie S4) during loading and unloading in the 1st cycle were captured using an infrared camera (Fig. 2C).

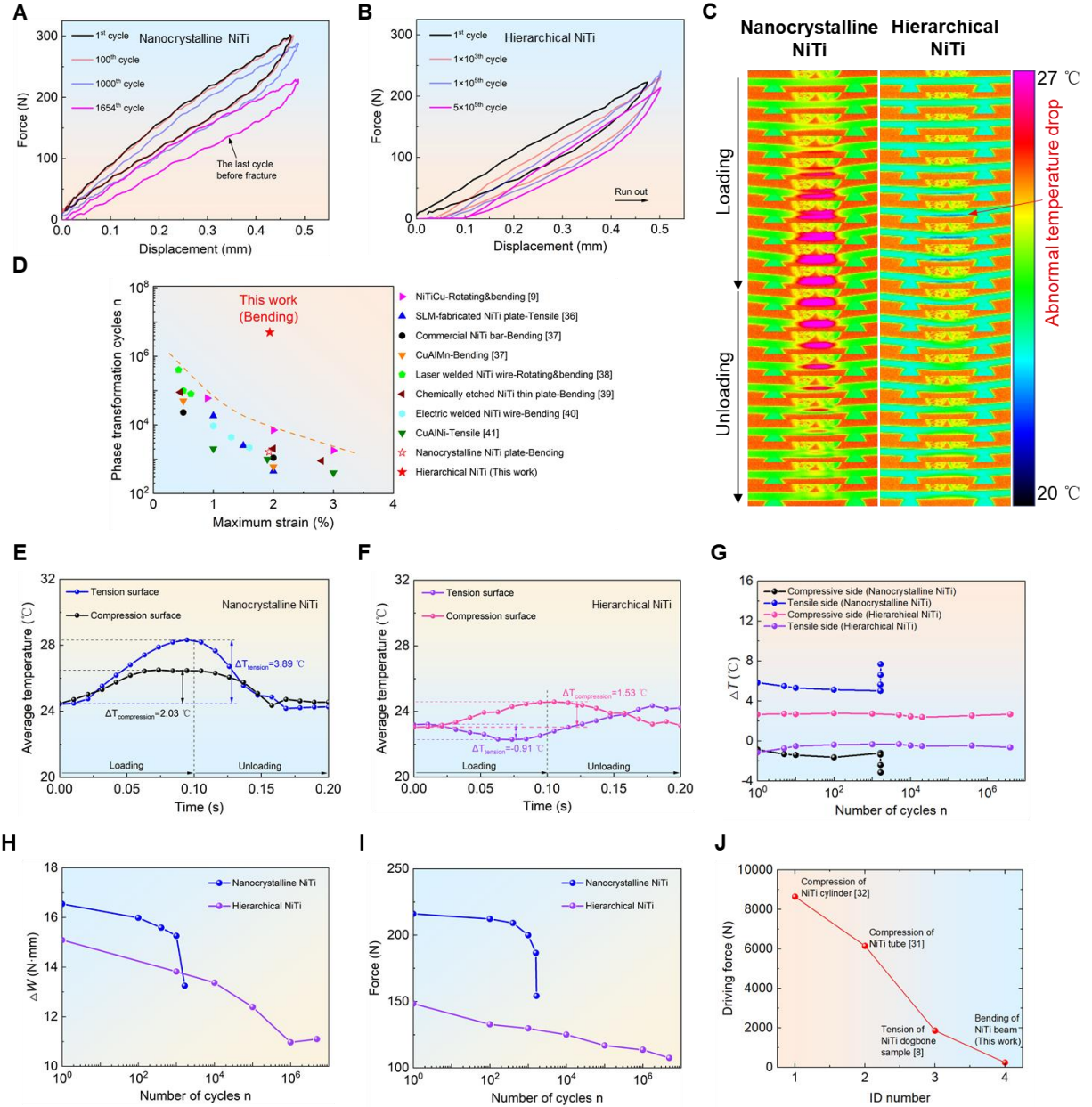

**Fig. 2. Cyclic bending cooling and fatigue behaviour of nanocrystalline and hierarchical NiTi SMA.** (**A-B**) Force-displacement curves of the nanocrystalline and hierarchical NiTi. (**C**) Infrared images of the nanocrystalline and hierarchical NiTi during loading and unloading. (**D**) Comparison of fatigue life and maximum tensile strain between this work and the values reported in literature. (**E-F**) Average temperature-time curves of the tension and compression

surfaces of the nanocrystalline and the hierarchical NiTi. (**G**) Evolution of $\Delta T$ (maximum temperature change) with the number of cycles. (**H-I**) Evolution of dissipated energy ($\Delta W$) and force at 0.3 mm displacement with the number of cycles. (**J**) Comparison of the maximum driving force under different loading modes.

A comparison of the fatigue life in our study with that reported in prior literature shows that the fatigue life of the hierarchical NiTi SMA in this study far exceeds the reported results in the literature (Fig. 2D) (*9, 31-36*). Achieving $5 \times 10^6$ bending cycles of the NiTi SMA under the maximum strain of 1.94% surpassed the required fatigue life ($3.9 \times 10^6$ cycles for a 10-year lifespan, accounting for half a day of six months per year, at a frequency of 0.05 Hz) of elastocaloric materials for refrigeration. The results stand as the highest values of bending fatigue life of NiTi SMA (under 1.94% tensile strain) reported to date.

The average temperature-time curves of the nanocrystalline and hierarchical samples in the 1$^{st}$ cycle were obtained from the infrared data (Fig. 2, E and F). Both nanocrystalline and hierarchical samples exhibit tension-compression asymmetry. In the nanocrystalline sample, both the tensile and compressive sides show a temperature increase during loading and a decrease during unloading. The tensile side demonstrates a temperature increase of 3.89 °C, while the compressive side demonstrates an increase of 2.03 °C during loading. However, in the hierarchical sample, the tensile side undergoes a small temperature decrease (0.91 °C) followed by an increase in temperature (0.38 °C) during loading, and a temperature rise (1.69 °C) followed by a decrease in temperature (0.18 °C) during unloading. On the compressive side, the temperature increases (1.53 °C) during loading and then decreases (1.53 °C) during unloading, similar to the as-received sample. The evolution of $\Delta T$ (maximum temperature change) with the number of cycles can be counted (Fig. 2G). For the nanocrystalline sample, $\Delta T$ remains relatively steady for the first 1629 cycles but undergoes a dramatic variation in the last 25 cycles, indicating sample failure in the final 25 cycles. In contrast, for the hierarchical sample, $\Delta T$ remains constant throughout all cycles, with compressive and tensile $\Delta T$ values approximately 2.67 and -0.34 °C, respectively.

The evolution of dissipated energy ($\Delta W$) and force at 0.3 mm displacement with the number of cycles can also be obtained. It is observed that the hierarchical structure dramatically retards the decrease in $\Delta W$ and force at 0.3 mm (Fig. 2H). The as-received sample experiences a rapid decline in $\Delta W$ from 15.09 mJ at the 1$^{st}$ cycle to 10.97 mJ at the 10$^{6th}$ cycle, and then maintains stability until the $5 \times 10^{6th}$ cycle. The force level gradually decreases in the first 1,000 cycles for the as-received sample and then drops dramatically, indicating rapid fatigue fracture propagation (Fig. 2I). In contrast, the pw-LSP-treated sample shows a linear decrease in force with increasing cycle number, indicating superior fatigue resistance. The bending-type elastocaloric cooling prototype using a NiTi beam requires the least driving force compared to other elastocaloric cooling methods for NiTi comparing with the literature (*8, 37, 38*) (Fig. 2J). This innovative bending-type loading approach opens up new possibilities for fatigue-resistant solid-state cooling with compact and cost-effective motors, addressing the challenges associated with high driving forces in traditional elastocaloric cooling methods.

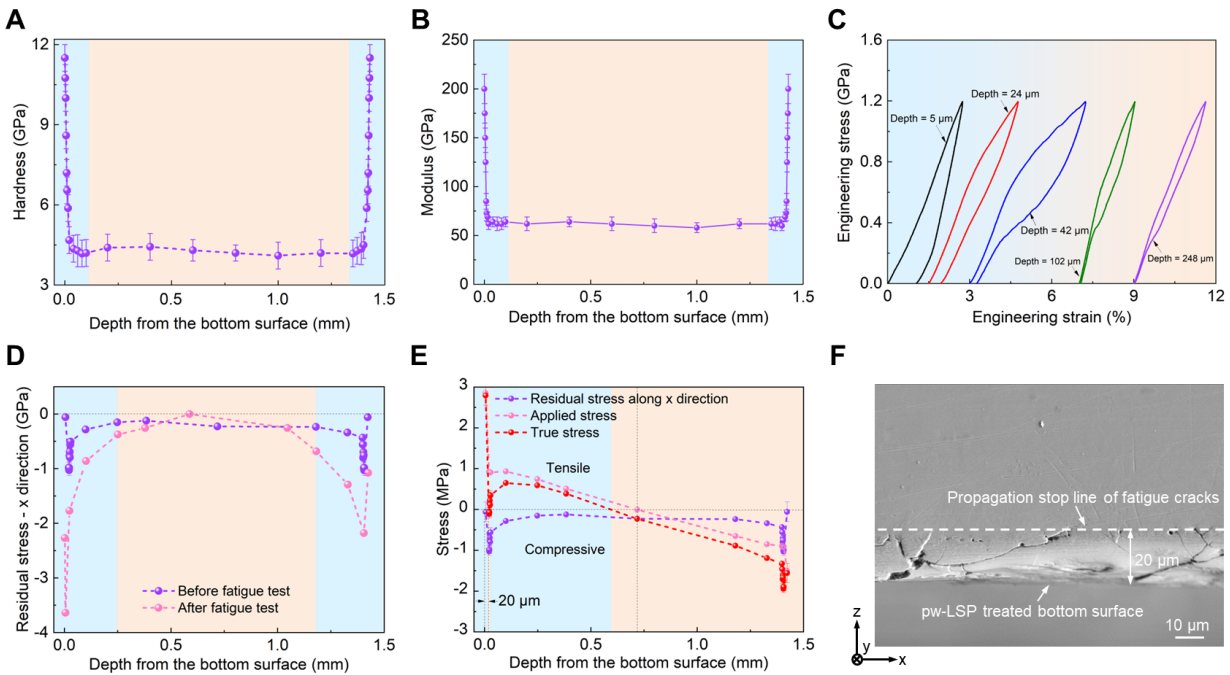

**Fig. 3. Characterization of the micromechanical behaviour of hierarchical NiTi SMA.** (**A**) Nano hardness distribution of different depths from the LSP-treated surface. (**B**) Young's modulus distribution of different depths from the LSP-treated surface. (**C**) Stress-strain curves of micropillars at different depths from the LSP-treated bottom surface. (**D**) Calculated residual stress along x direction before and after fatigue test at different depths from the LSP-treated bottom surface. (**E**) Residual stress along x direction, applied stress, and true stress at different depths from the LSP-treated bottom surface. (**F**) SEM image of the cross-section of the tensile surface after 5 million cycles.

To further explore the underlying mechanism responsible for the exceptional bending fatigue resistance, we conducted micromechanical tests at the cross-section of the specimen (fig.

S6). The load-displacement curves from the nanoindentation tests display a progressive increase in indentation depth from the surface to a depth of 20 μm, with the nanohardness (Fig. 3A) and elastic modulus (Fig. 3B) exhibiting a gradient distribution along the depth from the surface. Specifically, the measured nanohardness demonstrates an increase from 4.5 GPa at a depth of 20 μm to about 12 GPa at the surface and the elastic modulus shows a corresponding increase from 60 GPa to 200 GPa. The exceptionally high surface hardness can be attributed to the strengthening of precipitates including $Ni_4Ti_3$ and TiN (*39*). In addition, gradient distributions of the Ni/Ti ratio and element percentage at the top surface can be verified through energy dispersive X-ray (EDX) analysis (fig. S28).

To measure the residual stress at different depths, we first calculated the residual elastic strain across various depths using a focused ion beam-digital image correlation (FIB-DIC) method. This was accomplished by analyzing the mechanical behavior of cylindrical micropillars, each with a diameter (Φ) of 1 μm, milled at different depths from the pw-LSP-treated surface (fig. S6). The stress-strain curves obtained from these micropillars (Fig. 3C) reveal a noteworthy trend: as depth increases, the hysteresis loop area initially grows and then diminishes. This phenomenon arises from the non-monotonic relationship between the hysteresis loop area and the grain size in polycrystalline NiTi (*40*) . Furthermore, examination of the stress-strain curves of the micropillars at depths of 5 μm and 24 μm revealed a relatively large residual strain following compression of 1200 MPa. This observation suggests that the alloy exhibits partial superelasticity at these depths, attributed to the formation of R, B19′, and non-transformable phases (including TiN, and $Ni_4Ti_3$) at the surface treated with pw-LSP.

Evident gradient distributions of relief strain and residual elastic strain in the pw-LSP-treated sample are seen along the *x*- and *z*-axes (figs. S29 and S30). By integrating the stress-strain curves and residual elastic strain at different depths, we estimated the residual stress at various depths from the pw-LSP-treated surface before and after fatigue test (Fig. 3D and fig. S31). The residual stress along the x-direction was found to be more significant after the fatigue test, likely due to cyclic compression and stress redistribution to maintain balance within the specimen. As the fatigue cycle progresses, the residual strain increases while the applied strain decreases, leading to an expansion of the compressive region and an increase in compressive residual stress. This further provides a source for the substantially increased bending fatigue life of the pw-LSP-treated NiTi SMA. The residual stress along the *x*-direction ($\sigma_{res}$), applied strain ($\sigma_{app}$), and true strain ($\sigma_{true} = \sigma_{res} + \sigma_{app}$) at different depths from the pw-LSP-treated bottom surface can be calculated with Hooke's law (Fig. 3E). Notably, the minimum residual stress along the *x*-direction on the topmost surfaces can reach -1026 MPa. This indicates that at a tensile stress of 917 MPa on the same depth, there will be a region (at approximately 20 μm in depth) experiencing no tensile stress on the sample subsurface during the entire loading and unloading process. This phenomenon significantly enhances the fatigue life of the sample. Moreover, the stress intensity factor under compression loading is much lower than that of transverse fracture under cyclic tension (*38*). This further provides a source for the substantially increased bending fatigue life of the pw-LSP-treated NiTi SMA. Element mapping of the pw-LSP-treated surface after the fatigue test showed a uniform distribution (fig. S32). The fatigue fracture arrest after spreading to a depth of approximately 20 μm (Fig. 3F), possibly due to the compressive stress zone.

To further analyse the fatigue mechanism of the hierarchical NiTi SMA, *in situ* displacement-controlled cyclic tensile fatigue tests were conducted at a depth of ~13 μm from the pw-LSP-treated surface (Fig. 4, A to D). Analysing the engineering stress-strain curves of the micro-tension specimen at the 1st and 4000$^{th}$ cycle (Fig. 4A), we determined that the specimen displayed superelastic behaviour up to a maximum tensile strain of 6.9% and stress of 670 MPa. The maximum stress level decreased with increasing number of cycles, with the residual strain being ~4.8% after the 6650$^{th}$ cycle. The micro-tensile specimen after the 4,000$^{th}$ cycle reveal numerous traces on the surface of gauge length (Fig. 4C). These traces may result from the forward and reverse movement of phase interfaces during the cyclic tensile test, serving as evidence of phase transformation occurring during the tensile test. Engineering stress-strain curves under displacement-control at the 4000$^{th}$ to 4003$^{th}$ cycles show the residual strain to progressively increase with cycle number, with the specimen continuing to exhibit superelastic behaviour with a maximum engineering strain of 9.6% (fig. S33). However, with further increase in the number of cycles, microcracks were observed after 6650 cycles (Fig. 4D); these cracks were nearly perpendicular to the tensile direction, with occasional bifurcating cracks. Eventually, the specimen failed after 6651 cycles.

*In situ* displacement-controlled cyclic compressive fatigue tests were conducted at a depth of ~20 μm from the pw-LSP-treated surface (Fig. 4, E to H, fig. S34, and movie S5). By analysing the engineering stress-engineering strain curves of the micro-compression specimens at different cycles (Fig. 4E), we determined that the maximum stress level decreased with cycle number (fig. S35A), while the residual strain increased with cycle number (fig. S35B); specifically, the residual strain was ~4.5% after 10 million cycles (movie S6). Micro-voids were visible after 1 million cycles, with their size expanding with increasing cycles (Fig. 4, F to H).

TEM specimens were extracted at the fatigue crack tip of tensile side of the hierarchical NiTi beam after 5 million cycles (fig. S36), where it was seen that the upper unaffected region was crystalline, while the bottom shear band region was amorphous (Fig. 4I); this was further verified using HRTEM (Fig. 4J). The nature of the fracture process was that cracks propagated from the amorphous region to the crystalline region, where they tended to bifurcate, indicating that the crystalline regions act as obstacles to fatigue crack extension. Further observation of these crystalline regions near the crack tip indicated the existence of amorphous, R and B19′ phases (Fig. 4K), with a coherent interface between the R and B19′ phases (Fig. 4L). Our mechanistic analysis of the high bending fatigue resistance of the hierarchical NiTi is presented in Fig. 4M.

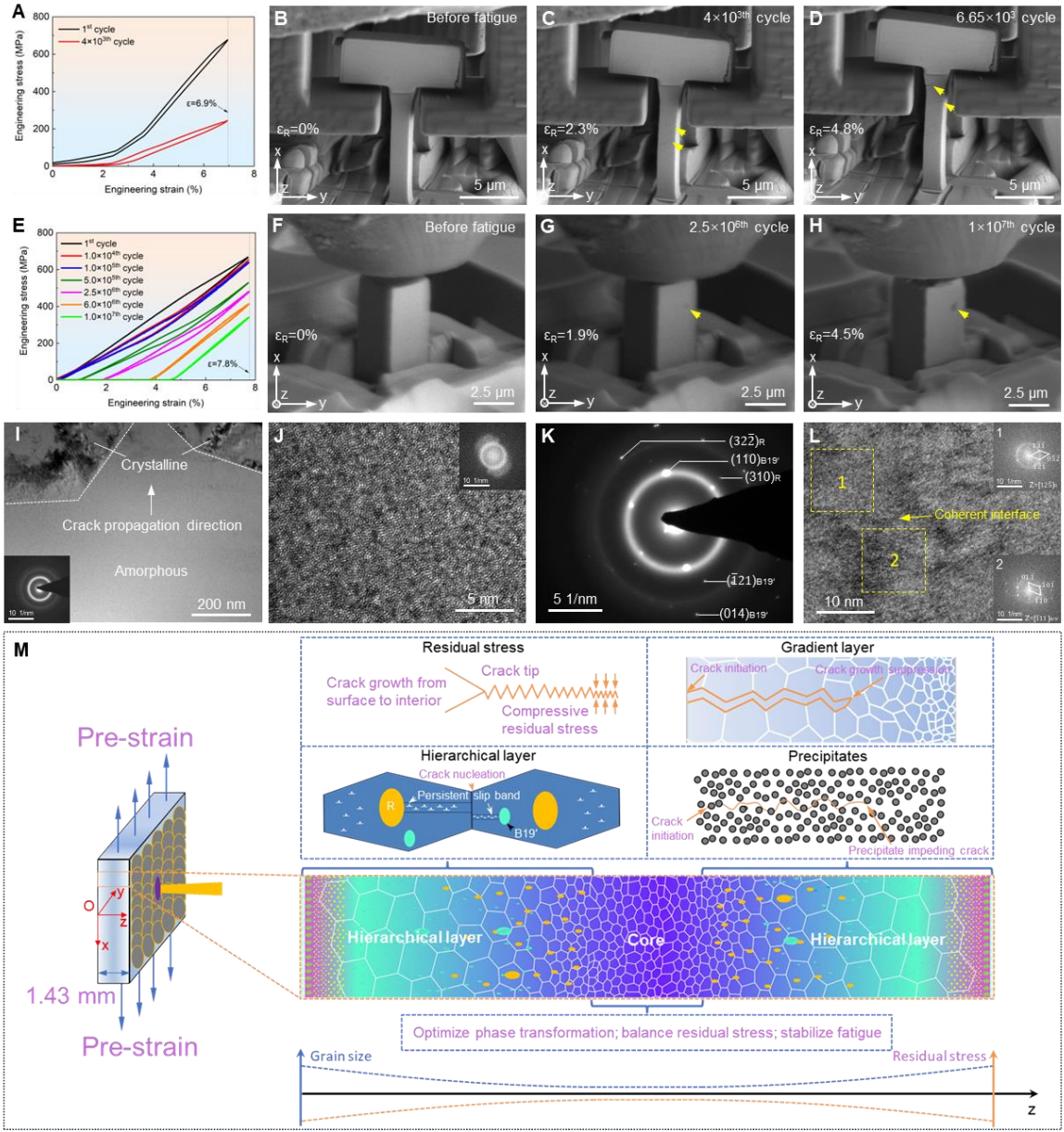

**Fig. 4. Fatigue mechanism of the hierarchical NiTi SMA.** (**A**) Engineering stress-strain curves of cyclic-tension specimen at the 1$^{st}$ and 4000$^{th}$ cycle. (**B-D**) *In situ* SEM images of the surface morphology of the cyclic-tension specimen at a depth of 13 μm. (**E**) Engineering stress-strain curves of cyclic-compression specimen from the 1$^{st}$ to 1×10$^{7th}$ cycle. (**F-H**) *In situ* SEM images of the surface morphology of the cyclic-compression specimen at a depth of 20 μm. (**I**) TEM image of the fatigue crack tip of hierarchical NiTi beam specimen after 5 million cycles bending, the inset is the SAED pattern of the middle region of (I). (**J**) HRTEM image of the middle region of (I), the inset is the FFT image of the middle region of (J). (**K**) SAED pattern of the crack tip region. (**L**) HRTEM image of the crystalline region near crack tip, showing the coherent

interface between R phase and B19′ phase, the insets are the FFT image of the selected regions. (**M**) Mechanism analysis of high bending fatigue resistance of hierarchical NiTi.

The processing temperature and pre-straining substantially influence the surface microstructure and macroscopic stress-strain response of the nanocrystalline NiTi SMA. Increasing the temperature from 25 °C to 150 °C can elevate the stress plateau of the as-received sample (fig. S37). Additionally, the dislocation density and cyclic stability can be considerably increased by raising the pre-strain (*41*). All these factors collectively contribute to a marked reduction in the Hugoniot elastic limit $\sigma_{HEL}$ of the NiTi SMA, thereby promoting plastic deformation. The surface hierarchical layer can inhibit the initiation of surface cracks, while the interior crystalline phase inhibits the development of fatigue cracks, given the higher crack growth resistance of the large grains compared to the small grains (*24*). The presence of nanoprecipitates changes the conventional B2↔B19′ phase transformation to R↔B2 reversed phase transformation during loading. As the B2↔R phase transition exhibits a much smaller transformation strain than the B2↔B19′ transition, this R-type transformation serves to enhance the cyclic stability (*42*). As the bending fatigue life of the NiTi SMA is predominantly governed by the regions under tensile loads, with a decrease in the surface tensile strain/stress, the $K_I$ (mode I stress intensity factor, which scales with the tensile stress) of the transverse fractures is significantly reduced. In particular, there is a thin layer (at a depth of ~ 20 μm) where no tensile stress is observed, even when the specimen is under tension. This forms a barrier for the propagation of fatigue cracks to the inner region of the specimen, effectively preventing structural failure.

In summary, we have successfully developed a fatigue-resistant NiTi SMA that is suitable for bending elastocaloric cooling through the introduction of a hierarchical microstructure and a large compressive residual stress at the surface region. We achieve a continuous maximum temperature drop of 0.91 K and a dissipated energy (Δ*W*) of 15.1 N·mm at a low specific driving force of 220 N. The fatigue life of NiTi SMA achieves over 5 million cycles under a maximum surface tensile strain of 1.94% for macroscopic cyclic bending. The incorporation of hierarchical layers acts to redistribute the stress at the specimen surface, which provides the superior crack growth resistance of the top gradient layer to effectively inhibit fatigue crack propagation, while preserving the fundamental superelastic behaviour in the subsurface region. Further, the presence of nanoprecipitates transforms the traditional B2↔B19′ phase transition into a B2↔R phase transition which increases the functional stability. Additionally, the specimen surface is subjected to a compressive residual stress of up to 1026 MPa, enabling the sub-surface layer to endure compressive stress during bending. The consequence of these synergistic effects of the hierarchical microstructure and the compressive residual stress is an exceptional improvement in the bending fatigue life of the NiTi plates by more than three thousand times through our pre-strained, warm laser shock peening, specifically from 1654 cycles before treatment to over 5 million cycles after the pw-LSP. These fatigue properties of this material significantly surpass those published in the literature for NiTi alloys, which may provide a solution to the limitation of NiTi under tensile loading. The combination of compressive residual stress and hierarchical microstructure not only preserves the inherent superelastic properties of SMAs but also substantially increases their bending fatigue life, paving the way for promising industrial applications for SMAs.

**Acknowledgments:**

The authors acknowledge the assistance of the SUSTech Core Research Facilities and Materials Characterization and Preparation Facility (MCPF) of HKUST.

**Funding:**

K.Y. and F.R. acknowledge the support of the National Natural Science Foundation of China (grants nos. 52122102 and 12302095). The authors from Southern University of Science and Technology (Sustech) are grateful for the support of the Basic and Applied Basic Research Foundation of Guangdong Province (grants nos. 2022B1515120082 and 2022A1515110908), and the Shenzhen Basic Research Program (grant no. JCYJ20220530113017040). Q.S. acknowledge the support of the Hong Kong Research Grant Council (grant no. 16212322), The Project of Hetao Shenzhen-Hong Kong Science and Technology Innovation Cooperation Zone (grant no. HZQB-KCZYB-2020083), and The Science, Technology and Innovation Commission of Shenzhen Municipality of China (grant no. SGDX20190816233605064).

**Author contributions:**

K.Y., F.R. and Q.S. conceived the idea and designed the research. K.Y., W.H. and Q.S. designed the processing parameters and prepared the specimens. K.Y. and W.H. operated the laser and manufactured the samples. K.Y. characterized the specimens by means of SEM, EDS, Nanoindentation, SPM and XRD. K.Y., K.C. and Q.Z. performed the TEM and STEM observations. K.Y. and P.W. performed the FIB-DIC measurement. K.Y., P.W. and K.C. conducted the fatigue testing. K.Y. conducted the FEM calculations. K.Y., K.C., P.H., P.W., H.G., Q.Z., F.R., Q.S., and R.O.R. analyzed the data and discussed the results. K.Y., K.C., P.H., P.W., H.G., Q.Z., W.H., F.R., Q.S. and R.O.R. wrote and revised the manuscript. All authors reviewed and contributed to the final manuscript.

**Competing interests:** The authors declare that they have no competing interests.

**Data and materials availability:** All data are available in the main text or the supplementary materials.

# Supplementary Materials for

# Low driving-force stable bending cooling via fatigue-resistant hierarchical shape memory alloy

Kai Yan[1,2], Kangjie Chu[1,2], Peng Hua[2,3], Pengbo Wei[1,2], Hanlin Gu[4], Qiming Zhuang[1], Weifeng He[5,6], Fuzeng Ren[1,*], Qingping Sun[2,3,*], Robert O. Ritchie[7,8,*]

*Corresponding authors. Email: renfz@sustech.edu.cn (F.R.); meqpsun@ust.hk (Q.S.); roritchie@lbl.gov (R.O.R)

**This PDF file includes:**

    Materials and Methods
    Supplementary Text
    Figs. S1 to S37
    Table S1 to S4
    References

## Materials and Methods

### Material fabrication

Contained from Johnson Matthey, USA, commercially available polycrystalline superelastic 50.8 at.% Ni-49.2 at.% Ti sheets, with an initial thickness of 1.5 mm, an austenite finish temperature of $A_f = 20.25$ °C (fig. S2), and an average grain size of 65 nm, were employed in this study (fig. S3). Dog-bone-shaped samples with a gauge length of 15 mm were wire-cut and polished using sandpaper, followed by additional $Al_2O_3$ suspension polishing with particle sizes from 1 μm, 300 nm, to 50 nm. After polishing, the ultimate thickness of the NiTi sheets was ~1.43 mm.

During the pre-strain warm laser shock peening (pw-LSP) process, a copper foil was placed on the processing surface to serve as a protective layer (fig. S1). Another NiTi plate with the same thickness was positioned below to prevent spallation caused by reflected tensile waves on the specimen rear surface. Additionally, a copper foil with a thickness of 0.2 mm was inserted between the NiTi sample and the shim. NiTi has a large stress plateau at room temperature, which makes this material very hard to endure plastic deformation at room temperature (fig. S4A). However, at high temperature, the material shows mechanical behaviour similar to traditional structural materials (fig. S4B), which is due to the higher transformation stress compared with the yield stress of austenite.

Prior to the LSP treatment, samples were pre-strained to 10%, ensuring the full conversion of the austenite phase to martensite. Subsequently, the sample was heated to 150 °C on a 2-mm-thick ceramic heating plate for 10 minutes. For the LSP treatment, a Q-switched Nd: YAG laser operating at 1 Hz with a wavelength of 1064 nm and a laser duration of approximately 20 ns was used to process the surface. The laser beam has a diameter of 2 mm and intensity of 7.9 GW/cm$^2$. Each pulse of the laser carried an energy of 4.97 J. The overlap ratio was maintained at approximately 50%. Each side of the NiTi plate underwent three rounds of LSP treatment, addressing both the top and bottom surfaces.

### Surface characterization

The surface morphology of the as-received (fig. S5A) and pw-LSP-treated (fig. S5B) NiTi shape memory alloy (SMA) was analysed to elucidate the mechanisms underlying the newly observed temperature drop under stress. Notably, the pw-LSP-treated surface revealed the presence of laser-induced periodic surface structure (LIPSS), numerous microcracks, and remelted particles. For further investigation, cross-sectional transmission electron microscope (TEM) samples were prepared at the pw-LSP-treated surface using the focused ion beam (FIB) lift-out method, as illustrated in fig. S5C. The scanning electron microscope (SEM) image of the thin foil distinctly exhibits a hierarchical microstructure in the pw-LSP-treated NiTi SMA (fig. S5D).

**<u>Microstructure characterization</u>**

The cross-sectional morphology and elemental distribution were examined using a JEOL-6390 scanning electron microscope (SEM) equipped with an energy-dispersive X-ray spectrometer (EDS). For the fabrication of transmission electron microscope (TEM) samples, an FEI Helios Nanolab 600i FIB-SEM dual beam system was employed. TEM characterization, including bright-field and dark-field TEM imaging, high-resolution transmission electron microscopy (HRTEM), and selected area electron diffraction (SAED), was carried out at 200 kV using a JEOL-2010 TEM. High-angle annular dark-field scanning transmission electron microscopy (HAADF-STEM) was performed using the FEI Talos TEM at 200 kV. The average grain sizes at various depths were calculated using multiple sets of dark-field TEM images, involving the counting of at least 500 grains with the assistance of the ImageJ program (ImageJ, National Institutes of Health, USA). X-ray diffraction (XRD) patterns were recorded on a PANalytical Empyrean X-ray diffractometer, utilizing Cu-Kα radiation with a step size of 0.1° and a count time of 1s per step.

Specimens for nanoindentation and residual stress measurements underwent mechanical polishing to achieve a mirror-like finish. Subsequently, electrochemical polishing using a mixture of 60% acetic acid and 40% perchloric acid was performed to remove any residual stress induced by mechanical polishing. Nanoindentation tests were conducted using a Hysitron Ti-950 Triboindenter equipped with a Berkovich diamond probe. The tests were performed under load-control with a loading rate of 200 μN/s and a peak load of 1000 μN, with a minimum of five indents carried out for each sample. The scanning probe microscopy (SPM) experiments were conducted using Hysitron Ti-950 Triboindenter under a scan rate of 0.1 Hz and a load amplitude of 2 μN.

**Residual stress measurement**

Residual stresses were measured using the FIB-DIC technique, as illustrated in fig. S6. Assuming the drilling depth is approximately equal to the circle's diameter, the residual stress of a circular component can be fully relieved based on the change in strain after drilling, facilitating the estimation of the relief strain and residual elastic strain at a specific depth. For the small strain region (residual elastic strain ≤ 1%), Hooke's law was applied to determine the residual stress distribution. This involved comparing the Young's modulus of measured sites, estimated from nanoindentations, with the calculated residual elastic strain obtained using the FIB-DIC method. In the case of the large strain region (residual elastic strain > 1%), the residual stress was determined by comparing the compressive stress-strain curves with the calculated residual elastic strain.

**Mechanical and fatigue testing**

Bending fatigue tests were carried out using a servo-hydraulic MTS 858 universal testing machine to assess the specimen's fatigue resistance both before and after pw-LSP treatment. A pre-load of approximately 15 N was applied to elevate the gap between the indenter and the specimen. Fatigue tests were conducted in a displacement-controlled mode with a maximum displacement of 0.5 mm and a loading frequency of 5 Hz (fig. S7). An infrared camera was employed to measure the *in situ* temperature field of the cross-section during cyclic testing (Fig. 2C). Theoretical calculations and finite element methods (FEM) were employed to calculate the maximum tensile and compressive strains on the specimen surfaces. High temperature tension tests were conducted with a universal tensile machine equipped with a high-temperature extensometer.

***In situ* micromechanical tensile experiment**

*In situ* uniaxial tensile tests were conducted on an FIB-milled T-shaped micro-slat using a FemtoTools Nanomechanical Testing System (model FT-NMT04, Buchs ZH, Switzerland) within a TESCAN MIRA3 SEM. The loading direction was aligned with the x-axis. Initially, a nanoindentation test was performed on the surface to calibrate the system compliance, essential for analyzing force–displacement data and eliminating any artificial deformations stemming from the grip and loading system. Subsequently, the tensile grip was positioned to secure the micro-slat with appropriate alignment adjustments. Tensile testing commenced with incremental increases in maximum displacement from 0 to 0.5 μm, employing a relatively rapid loading rate. Displacement–force data were collected under displacement control at a fixed rate of 76 nm/sec to capture the quasi-static mechanical responses of the tensile micro-slat while minimizing the impact of thermal drift as much as possible. Following this, cyclic fatigue tests were carried out under displacement control at a loading frequency of 5 Hz. SEM images of the T-shaped micro-slat were taken after several specific cycles.

**FEM analysis**

For the pw-LSP treatment process, a Johnson-Cook model was implemented to simulate the mechanical responses of the as-received NiTi specimen under pw-LSP; the detailed material parameters can be found in Table S3 and the Supplementary Text section. The input values were obtained from experimental results and the MatWeb Online Materials Database (www.matweb.com). The FEM simulation involved a 2-D model with dimensions of 3 mm in length and 1.43 mm in thickness. Boundary conditions included constraining displacement along the *y*-direction ($U_y$) of the bottom surface, and rotational degrees of freedom around the *z*-axis ($UR_z$) of the bottom and left surfaces. The 2-D finite element model was used for the simulation, employing coupled temperature-displacement elements with a mesh size of 0.050 mm. The specimen was subjected to a pre-defined internal stress of 1100 MPa along the *x*-direction, and a temperature field of 423 K was applied before external laser ablation. The convective heat transfer coefficient of the model was set at 0.01 W/(m$^2$•K) for the upper and bottom surface, and 1 W/(m$^2$•K) for the left and right surface. The laser ablation field was represented by a pressure of 70 MPa and a heat flux of $7.11 \times 10^{11}$ W/m², assuming a laser energy absorption percentage of 0.9.

For the bending deformation of the NiTi beams, a linear-elastic model was employed, incorporating a Young's modulus of 60 GPa and a Poisson's ratio of 0.30, to simulate the 4-point bending mechanical behaviour of the as-received NiTi specimen. For the pw-LSP-treated specimen, a gradient distribution of Young's modulus was applied, using the values derived from nanoindentation tests (Fig. 3B). The 2-D model was performed in ABAQUS using CPS4R elements, with an overall mesh size of ~0.05 mm. The simulation began by modelling the displacement induced by the pre-load, followed by calculating the strain field resulting from the displacement of the upper indenter.

## Supplementary Text

### Mechanism analysis of abnormal temperature drop

The mechanistic competition between the thermoelastic effect, inverse elastocaloric effect (R→B2 reversed-phase transition), and elastocaloric effect (B2→R forward phase transition) can be considered to provide the source of the unusual temperature variation (*43*). During loading, the axial stress $\sigma_{xx}$ of the hierarchical NiTi at ~20 μm depth stays compressive during the whole process (fig. S25C, $\sigma_{xx}$ increases from -1200 MPa to -220 MPa at the tensile surface during loading). Reverse phase transition primarily occurs at the tensile surface of the sample (temperature decreases during loading), and forward-phase transition primarily takes place at the compressive side of the specimen (temperature increases during loading). Conversely for nanocrystalline NiTi, the tensile surface primarily experiences forward phase transition with a high tensile stress ($\sigma_{xx}$ increases from 200 MPa to 1100 MPa). The R phase transformation results from the coherent and semi-coherent $Ni_4Ti_3$ precipitates. $Ni_4Ti_3$ precipitates smaller than 100 nm are coherent with the B2 austenite matrix, while larger $Ni_4Ti_3$ precipitates gradually lose coherency with increasing size (*44*). Since the size of $Ni_4Ti_3$ is small (<100 nm), the interface between the $Ni_4Ti_3$ and the B2 phase remains coherent, mitigating stress concentration at the phase interface, and thus preventing the occurrence of fatigue fracture. The strain variation due to the B2↔R phase transition is much smaller than that of B2↔B19′ phase transition, which contributes to the enhancement of fatigue life (*42*).

The hierarchical structure further contributes to the ultimate fatigue life enhancement. The laser affected layer extends approximately to 40 μm on each side, accounting for about 5.59% of the sample thickness. Processing both sides introduces hierarchical layers, redistributing stress at the specimen surface. The top region of the gradient layer impedes fatigue crack propagation by inducing higher crack-growth resistance (*24*), while the subsurface region maintains superelastic behavior for cyclic loading. Moreover, there exists a thin layer (at a depth of ~20 μm) that experiences a very small stress level throughout the entire bending fatigue test (figs. S22 to S25); this prevents the propagation of the fatigue fracture toward the inner depth, and thus dramatically enhances the fatigue lifetime. Consequently, the fatigue life of the samples processed with pw-LSP surpasses that of the as-received samples by a considerable margin. Fig. S26A shows the XRD patterns of the pw-LSP-treated sample surface (tensile side) before and after cyclic bending. Distinct diffraction peaks of the B2, R, and B19′ phases, as well as TiN, Ti$_2$N, TiO, and precipitates (including Ni$_4$Ti$_3$, Ti$_2$N, and Ni$_{2.67}$Ti$_{1.33}$) can be identified. The transformation path shifts from B2↔B19′ into B2↔R↔B19′ in NiTi SMAs when Ni$_4$Ti$_3$ precipitates are present in the B2 matrix. The peak intensity of the R-phase, at approximately 36.5°, 44.2° and 73.9°, increases with the cycle number, indicating B2↔R transformation and the accumulation of residual R phase during cyclic bending. The computed full-width-at-half-maximum (FWHM) of the (110) B2 progressively decreases as the number of cycles increases (fig. S26B), probably due to the accumulation of dislocations (*45*) and residual stress (*46*) during the bending fatigue test.

**Calculation of residual elastic strain and relief strain**

The residual elastic strain (*47*) along the x and z directions (defined in fig. S6) can be calculated using Eqns. (1-2).

$$\varepsilon_{rxx} = -\frac{(\varepsilon_{xx} + \nu\varepsilon_{zz})}{(1-\nu^2)} \quad , \tag{1}$$

$$\varepsilon_{rzz} = -\frac{(\varepsilon_{zz} + \nu\varepsilon_{xx})}{(1-\nu^2)} \quad , \tag{2}$$

where $\varepsilon_{xx}$ and $\varepsilon_{zz}$ respectively represent the relief strain along the x and z directions, $\nu$ is Poisson's ratio, and $\varepsilon_{rxx}$ and $\varepsilon_{rzz}$ are respectively the residual elastic strain along the x and z directions. By measuring the change in distance of two points along the x and z directions, $\varepsilon_{rxx}$ and $\varepsilon_{rzz}$ can be calculated using Eqns. (3-4).

$$\varepsilon_{xx} = \frac{L_{x1} - L_{x0}}{L_{x0}} \quad , \tag{3}$$

$$\varepsilon_{zz} = \frac{L_{z1} - L_{z0}}{L_{z0}} \quad , \tag{4}$$

where $L_{x0}$ and $L_{z0}$ represent the actual distance between two marks and $L_{x1}$, and $L_{z1}$ the distance of the two marks after stress relaxation by drilling a rectangular trench.

**Calculation of the Hugoniot stress limit**

As described by Fabbro's model (*48*), the peak shock pressure $P$ as a function of laser intensity $I_0$ can be expressed as:

$$P(I_0) = 0.01\sqrt{\frac{\alpha}{2\alpha+3} \cdot Z \cdot I_0} \quad , \tag{5}$$

$$\frac{2}{Z} = \frac{1}{Z_{air}} + \frac{1}{Z_{NiTi}} \quad , \tag{6}$$

$$Z_{air} = \rho_{air} D_{air} \quad , \tag{7}$$

$$Z_{NiTi} = \rho_{NiTi} D_{NiTi} \quad , \tag{8}$$

where $\alpha$ is the portion of absorbed energy contributed to the thermal energy of plasma and is taken as 0.25 in this study (*49*), $Z_{air}$ and $Z_{NiTi}$ are the shock impedance of the target material and the confining media determined by the material density $\rho$ and shock velocity $D$. Specifically, as $\rho_{air} = 0.00129$ g/cm$^3$, $\rho_{NiTi} = 6.5$ g/cm$^3$, $D_{air} = 340$ m/s, $D_{NiTi} = 5150$ m/s, then $Z_{air} = 43.86$ g cm$^2$ s$^{-1}$ and $Z_{NiTi} = 3.35 \times 10^6$ g cm$^2$ s$^{-1}$. From Eqn. (6), we can then calculate $Z = 87.72$ g cm$^2$ s$^{-1}$. Considering the values of each parameter, Eqn. (5) can be simplified to:

$$P_{laser} = 0.025\sqrt{I_0} \quad . \tag{9}$$

Using these calculation, we can determine that the peak pressure is ~70 MPa for a laser intensity of 7.9 GW/cm$^2$.

Ballard (*50*) has developed an analytical model for calculating the residual stress induced by LSP. Using this Ballard model, one can determine the following:

(i) Optimum shock conditions for a given elastic-plastic material are defined by $\sigma_Y$, $E$, and the anisotropy coefficient.

(ii) To assess the fields of plastic deformation and residual stress caused by an impact pressure "$P$", this model assumes that (a) shock waves propagating in a perfect elastic-plastic half-space are longitudinal and planar, (b) the plastic strain follows a von Mises plasticity criterion, and (c) the applied pressure on the affected region is uniform.

The Hugoniot stress limit, $\sigma_{HEL}$, is defined by:

$$\sigma_{HEL} = \left(1 + \frac{\lambda}{2\mu}\right)(\sigma_Y - \sigma_0) = \frac{1-\nu}{1-2\nu}(\sigma_Y - \sigma_0) \approx 2(\sigma_Y - \sigma_0) \quad , \tag{10}$$

where $\sigma_Y$ is the compressive static yield strength, $\sigma_0$ is the internal stress, $\lambda$ and $\mu$ are the Lamé coefficients, $\lambda = \frac{E\nu}{(1-\nu)(1-2\nu)}$, and $\mu = G = \frac{E}{2(1+\nu)}$.

As the phase has entirely transitioned into martensite (10% pre-strain) and the yield stress of the martensite decreases considerably with temperature, the heat impact of the laser will induce martensite to transform back into austenite. As a result of the inherent temperature dependency of the stress (Claperyon–Clausius relation) (*51*), the estimated stress level will grow from 420 MPa to approximately 1300 MPa, which will cause $\sigma_{HEL}$ to continue to fall. As the peak pressure decreases during the propagation of a stress wave, there will be no plastic deformation if the peak pressure is less than $2\sigma_Y$, allowing for a hierarchical layer on the surface of an LSP-treated sample. Consequently, Eqn. (10) explains three processes for creating a hierarchical using the pw-LSP treatment:

(i) above $2\sigma_{HEL}$, the reverse elastic strain saturates to $4\sigma_Y$, and plastic strain occurs;

(ii) between $\sigma_{HEL}$ and $2\sigma_{HEL}$, plastic strain occurs with a purely elastic reverse strain;

(iii) below $\sigma_{HEL}$, no plastic deformation occurs.

Eqn. (10) shows that with the increase in $\sigma_0$, $\sigma_{HEL}$ will decrease by a factor of two. For example, a pre-stress of 400 MPa can decrease $\sigma_{HEL}$ by approximately 800 MPa. This may explain why pre-strain or prestress can enhance the effect of LSP.

**Calculation of the effect of LSP**

The depth of the residual stress layer ($L_p$), the plastic strain of the pw-LSP-treated surface ($\varepsilon_p$), and the residual stress of the pw-LSP-treated surface ($\sigma_{surf}$) can also be determined by:

$$L_p = \left(\frac{C_{el}C_{pl}\tau}{C_{el}-C_{pl}}\right)\left(\frac{P-\sigma_{HEL}}{2\sigma_{HEL}}\right) \quad , \tag{11}$$

$$\varepsilon_p = \frac{-2\sigma_{HEL}}{3\lambda+2\mu}\left(\frac{P}{\sigma_{HEL}}-1\right) \quad , \tag{12}$$

$$\sigma_{surf} = \sigma_0 - \left[\frac{\mu\varepsilon_p(1+\nu)}{(1-\nu)} + \sigma_0\right]\left[1 - \frac{4\sqrt{2}}{\pi}(1+\nu)\frac{L_p}{a}\right] \quad , \tag{13}$$

where $C_{el}$ and $C_{pl}$ are the velocities of elastic and plastic wave, $\tau$ is the duration of a pressure pulse, $P$ is the pressure, $\lambda$ and $\mu$ are Lamé constants, $a$ is the edge length of a square laser spot, and $\sigma_0$ is the initial surface residual stress (*52*).

From the above equations, the depth of the residual stress layer can be found to increase with an increase in the peak pressure and a decrease in $\sigma_{HEL}$; thus, increasing the laser power density and increasing the pre-strain will both increase the residual stress affected layer. From Eqn. (12), it is evident that if $\sigma_{HEL} < P$, $\varepsilon_p > 0$, while if $\sigma_{HEL} > P$, $\varepsilon_p < 0$.

## Calculation of the surface heat flux induced by laser ablation

Laser ablation is a process during which the material is heated above its boiling temperature very rapidly. To model this rapid heating process, the absorptivity is assumed to be $\eta = 90\%$. The absorbed energy ($E_{abs}$) can be calculated as: $E_{abs} = E * \eta = 4.97 * 90\% = 4.473$ J.

For the nanosecond laser shock peening treatment of metal in air, as the energy for one pulse of laser is 4.97 J, the pulse width of nanosecond laser ($\Delta t$) is 20 ns, and the diameter of laser beam is 2 mm, the surface heat flux ($q$) can be calculated as:

$$q = E_{abs} / (S * \Delta t), \quad (14)$$

where $S$ is the laser beam area and $\Delta t$ is the laser pulse.

From Eqn. (14), the calculated surface heat flux is 7.11 GW/cm$^2$.

## Theoretical calculation of the surface temperature induced by laser ablation

Laser ablation is a process during which the material is heated above its boiling temperature very rapidly. To model this rapid heating process, the absorptivity Å is assumed to depend on the temperature according to the following expression (*53*):

$$Å(T(r,0,t)) = a_0 \frac{T(r,0,t)}{T_v} \text{ for } T < T_v, \quad (15)$$

$$Å(T(r,0,t)) = a_1 (1 + \exp(a_2 \frac{T_v}{T(r,0,t)})) \text{ for } T > T_v. \quad (16)$$

where $T_v$ denotes the vaporization temperature of material. Eq. (15) describes a linear increase with temperature until the surface temperature reaches the vaporization temperature. The authors assumed an exponential decrease of the surface absorptivity for temperatures $T > T_v$ to account for the laser beam attenuation in the plasma plume. Parametric studies were conducted by taking $a_0 = 0.4$; $a_1 = 5.41$ and $a_2 = 10^{-5}$ in Eqs. (15) and (16). Defining $T^*(r, z, t) = T(r, z, t)-T_0$ and assuming that the problem is symmetrical, the unsteady heat conduction equation and the boundary and initial conditions in cylindrical coordinates can be written as

$$\frac{1\partial}{r\partial r}\left(r\frac{\partial T^*}{\partial r}\right) + \frac{\partial^2 T^*}{\partial z^2} = \frac{1\partial T^*}{\alpha \partial t} \quad . \tag{17}$$

and

$$T^*(r = \infty, z, t) = 0 \quad . \tag{18}$$

$$T^*(r, z = \infty, t) = 0 \quad . \tag{19}$$

$$-k\frac{\partial T^*(r,z=0,t)}{\partial z} = \text{Å}(T(r,z=0,t))I\exp\left(-2\frac{r^2}{r_0^2}\right)F(t) \quad . \tag{20}$$

$$T^*(r, z, t = 0) = 0 \quad . \tag{21}$$

$T^*(r = 0, z, t)$ must be infinite. (22)

Using the Hankel and Laplace transforms, we can simplify the pulse shape as rectangular shape for simplification, and the temperature can be calculated as:

$$T^*(r,z,t) = \frac{I_0 r_0^2 \sqrt{\alpha}}{k\sqrt{\pi}} \int_0^t \frac{\text{Å}^*(t-t')F(t-t')\exp\left(-\left(\frac{2r^2}{r_0^2}\right)-(z^2/4\alpha t')\right)}{\sqrt{t}\quad r_0^2+8\alpha t'} dt'' \quad . \tag{23}$$

For the operating conditions considered in this investigation ($r_0$=1 mm, $\alpha\sim10^{-4}$ m$^2$s$^{-1}$ $\tau\sim$30 ns), it can be shown that $r_0^2 + 8\alpha t' \approx r_0^2$. The singularity at $t'$ is removed by substituting $t'=t''/$t.

**Two-temperature model**

The two-temperature model is commonly used to describe the interaction of ultrafast laser with materials and to understand how ablation depth and effective penetration depth depend on pulse duration (*54*). This model operates on the assumption that the energy absorbed by the material is divided into two subsystems: electrons and phonons. Fast local thermalization of electrons occurs when the laser excites electrons, followed by thermalization between electrons and phonons. The lattice temperature and the electron temperature in the two-temperature equation are defined by the following two partial differential equations:

$$H_e \frac{\partial T_e}{\partial t} = \nabla(K_e \nabla T_e) - G(T_e - T_l) + Q(r,t) \quad . \tag{24}$$

$$H_l \frac{\partial T_l}{\partial t} = \nabla(K_l \nabla T_l) + G(T_e - T_l) \quad . \tag{25}$$

where the subscripts $e$ and $l$ in the differential equation denote electrons and lattices, respectively, while $t$ represents time, $H$ is the heat capacity, and $k$ is thermal conductivity. $G$ stands for the electron-lattice coupling factor, and $Q$ stands for the absorbed laser heating source.

**Johnson-Cook model of NiTi and copper**

The Johnson-Cook model (*55*) is a semi-empirical model which includes 5 material constants, and takes into account strain rate, strain hardening, and thermal softening effect. It can be written as:

$$\sigma = (A + B\varepsilon_p^N)\left(1 + C\ln\frac{\dot{\varepsilon}}{\dot{\varepsilon}_0}\right)[1 - (\frac{T-T_r}{T_m-T_r})^M] \quad , \tag{26}$$

where $\sigma$ is yield stress, $\varepsilon_p$ is effective strain of the plastic deformation, $\dot{\varepsilon}$ is effective strain rate of the plastic deformation, $\dot{\varepsilon}_0$ is the reference strain rate, and $T$ is the material temperature. $T_r$ and $T_m$ are the reference temperature and material melting temperature respectively. *A, B, N, C, M* are parameters determined from experimental data.

While quasi-isentropic compression experiments can yield information on strain hardening, the parameters deduced from them are subject to considerable uncertainty. It is important to note that in our experiments, the effective strain resulting from plastic deformations is relatively small (approximately 0.1 for isentropic compression experiments and 0.2 for shock experiments). Consequently, the strain hardening terms in the model can be safely disregarded. This allows for the application of a simplified Johnson-Cook model to describe the yield stress of the austenitic NiTi alloy:

$$\sigma = A\left(1 + C\ln\frac{\dot{\varepsilon}}{\dot{\varepsilon}_0}\right)[1 - (\frac{T-T_r}{T_m-T_r})^M] \quad . \tag{27}$$

The detailed material properties in Eqn. (27) are illustrated in Table S4.

**Numerical calculation of the surface temperature induced by laser ablation**

Given the ultrafast and complex nature of the pw-LSP process, theoretical calculation or experimental measurement of stress and temperature variations is challenging. Therefore, FEM simulations were employed to simulate the pw-LSP process (fig. S17).

Due to the plasma explosion at the surface, the surface velocity changes with time. According to the previous boundary condition, we can get the $\sigma_{zz}$ (stress along the $z$-direction)-time, and magnitude of the heat flux over area-time curves of the near-surface region (fig. S18).

The strain rate during pw-LSP was calculated; it was found that the strain rate varies during pw-LSP treatment, the minimum strain rate can reach about $-3.72 \times 10^4$ s$^{-1}$ along the $x$-direction, and about $-1.67 \times 10^6$ s$^{-1}$ along the $z$-direction (fig. S19). The strain rate is observed to fluctuate with time, likely due to the reflection of stress waves between the interfaces of different materials during propagation (fig. S20).

During LSP processing, the plasma temperature can reach 10,000 °C, as noted by Peyre *et al*. (*56*). This high temperature is highlighted as a contributing factor to the formation of the hierarchical microstructure and the generation of residual stress at the top surface region of the sample. There are mainly two kinds of models to calculate the surface temperature induced by the laser: a macroscopic model (*53*) and a two-temperature model (*54*). The macroscopic model is more suitable for nanosecond laser shock peening. However, it is relatively hard to get analytical solutions for this model, but from the FEM simulation results, the temperature at different time and depths from the pw-LSP-treated surface can be calculated (fig. S21). It is found that the temperature at the topmost region has a sudden increase then decreases dramatically (fig. S21A), and the highest temperature decreases with the increase of depth. From the temperature field of the specimen at $3.7 \times 10^{-5}$ s after LSP treatment, it is seen the thickness of the heat-affected zone is approximately 50 μm (fig. S21B). The gradient temperature profile at $3.7 \times 10^{-5}$ s (fig. S21C) shows that the highest surface temperature is approximately $8.8 \times 10^4$ K and decreases with depth. The surface temperature of a substance treated by pw-LSP dropped rapidly to room temperature. As a result of convection and slow heat conduction, the interior temperature decreases gradually. Therefore, the potential for grain growth at the pw-LSP-treated surface is inhibited. In contrast, grains in the inner layer are exposed to high temperatures for a longer time period and thus grain growth was induced. With the increase in depth, the heat effect of pw-LSP decreases gradually; this explains the formation of the gradient microstructure. At the topmost region, the ultra-high temperature will induce remelting of the NiTi alloy, such that the alloy will have reaction with the nitrogen and oxygen in air (*57*).

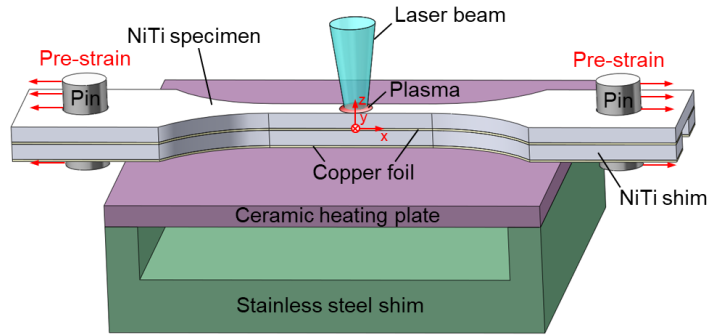

**Fig. S1.**
**Schematic illustration of the pw-LSP processing.** The specimen is pre-strained to 10% and heated to 150 ℃ before laser shock peening treatment.

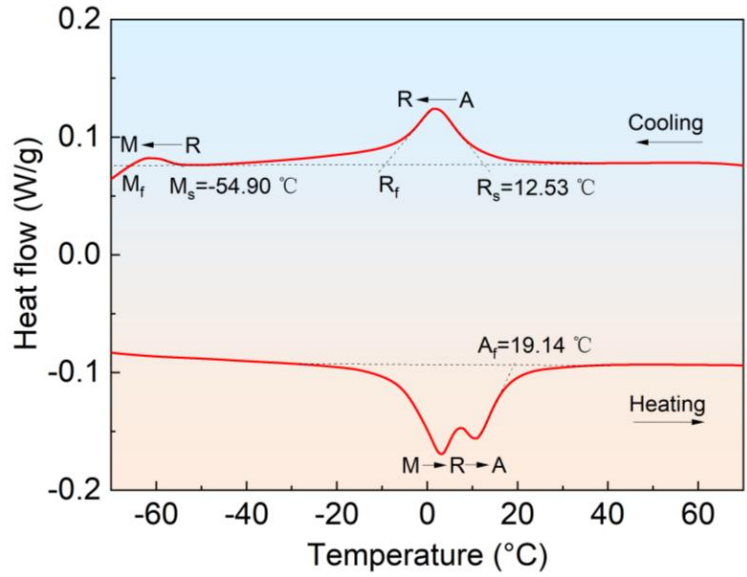

**Fig. S2.**
**DSC curve of the as-received nanocrystalline NiTi specimen.** It is seen that the austenite finish temperature of the as received specimen is about 20.25 °C.

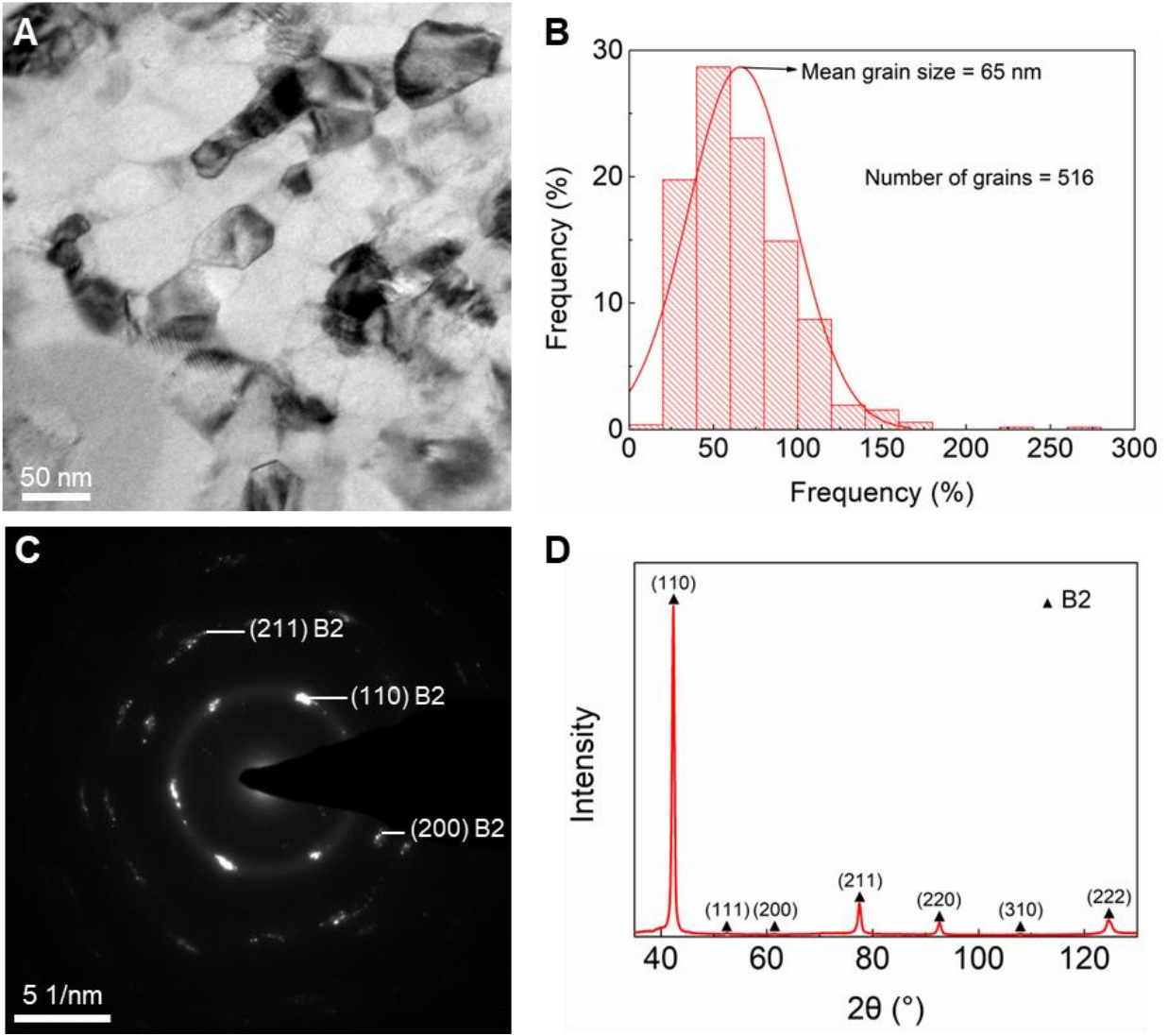

**Fig. S3.**
**Microstructure of the as-received NiTi sample.** (**A**) Bright-field TEM image. (**B**) The histogram of grain size distribution. (**C**) The corresponding SAED pattern of (A). (**D**) XRD pattern.

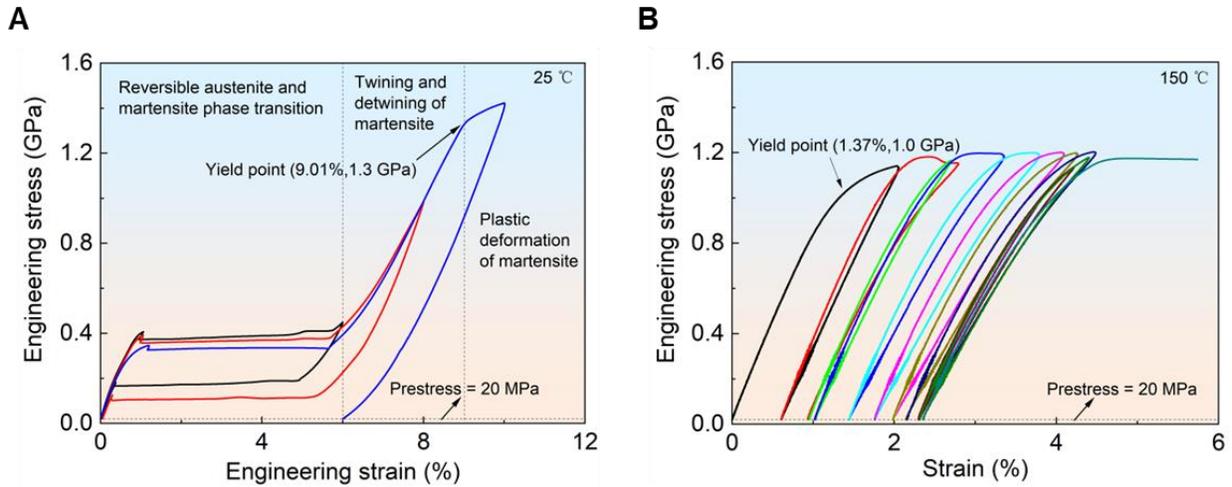

**Fig. S4.**
**Cyclic tensile engineering stress-strain curves of the as-received NiTi at 25 ℃ (A) and 150 ℃ (B).** At 25 ℃, the material shows superelastic behaviour. The yield strain of martensite is about 9.01%, with a corresponding yield strength of around 1.3 GPa. In contrast, at 150 ℃ the yield behaviour of austenite resembles that of traditional structural materials, with a yield strain of about 1.37% and a yield strength of approximately 1.0 GPa.

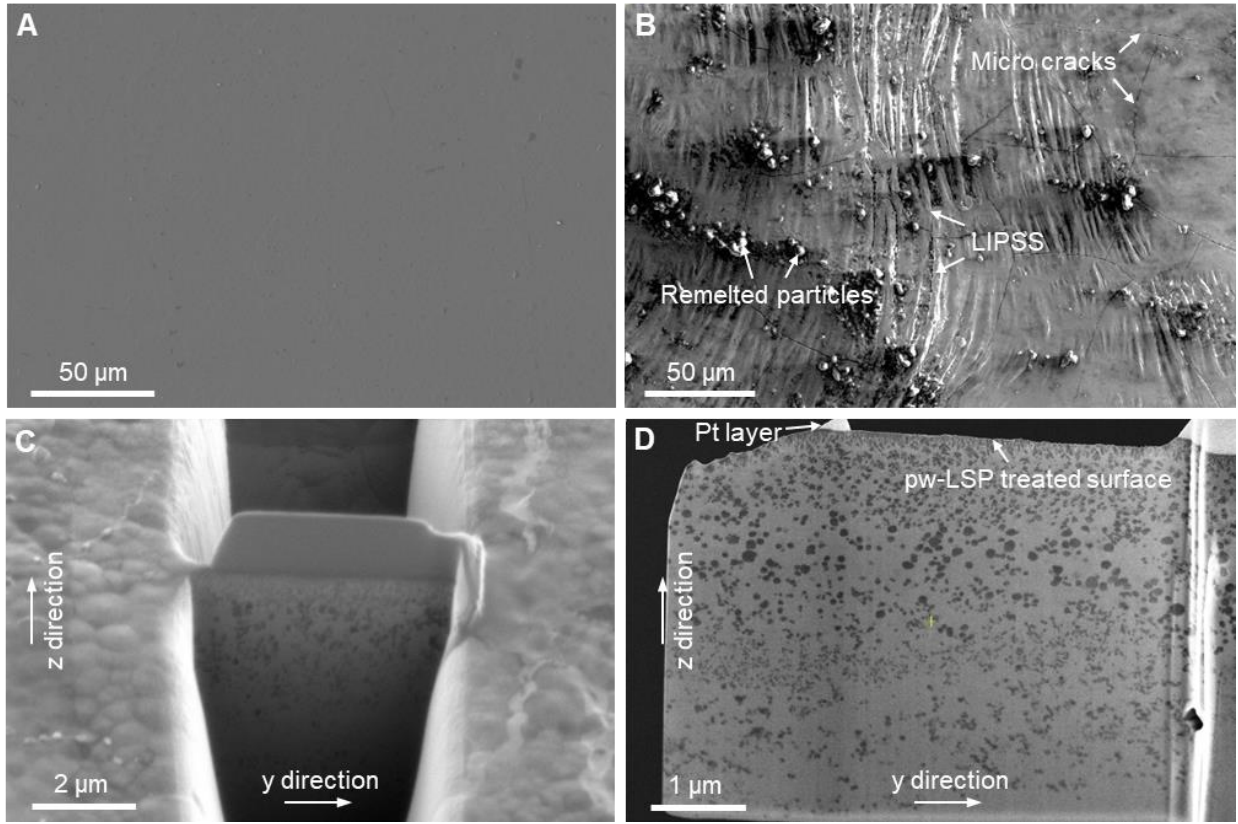

**Fig. S5.**
**Surface morphology and cross-sectional microstructure of the TEM sample**. (**A**) SEM image of the as-received NiTi specimen surface. (**B**) SEM image of the pw-LSP-treated surface. (**C**) TEM sample preparation using FIB lift-out technique. (**D**) SEM image of the cross-sectional TEM thin foil lifted out from the pw-LSP-treated surface region.

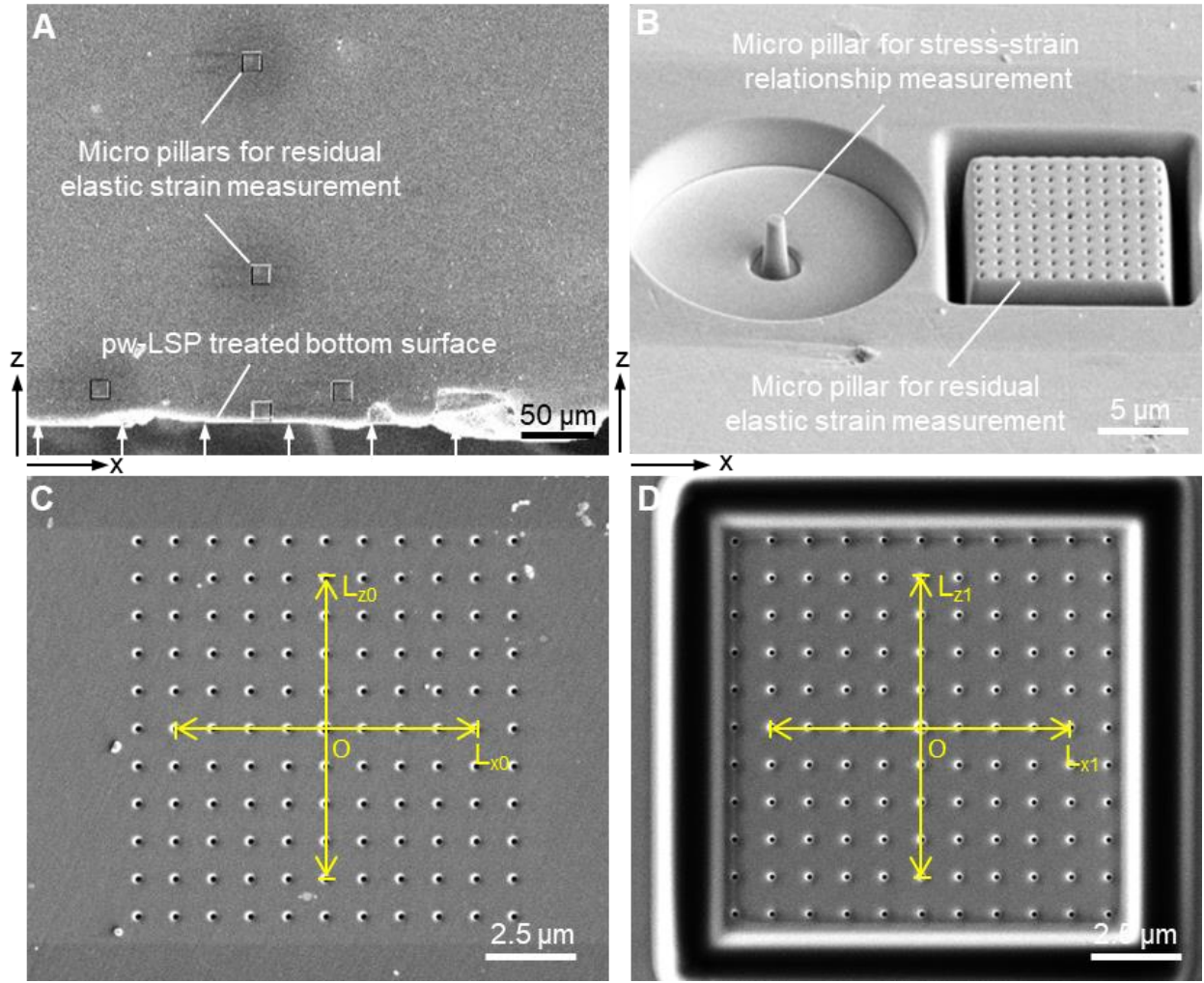

**Fig. S6.**
**Methodology of residual stress measurement with FIB-DIC method.** (**A**) SEM image of the cross-section of the pw-LSP-treated sample after FIB-DIC measurement. (**B**) SEM image of micropillars for the compression test and relief strain measurement. (**C-D**) SEM images of the dot matrix before and after the residual stress relaxation by milling a rectangular trench for residual stress measurement.

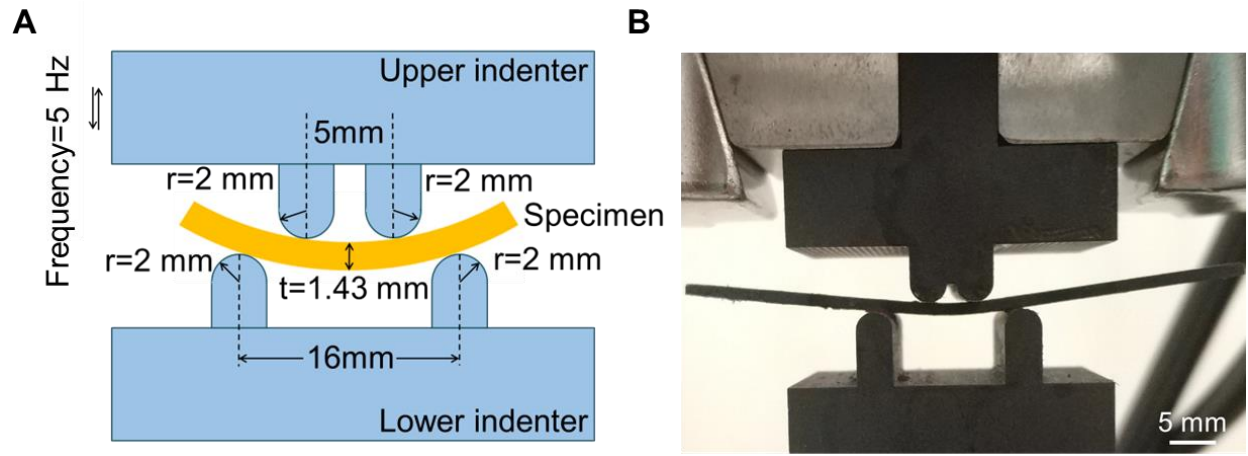

**Fig. S7.**
**Schematic diagram (A) and photograph (B) of the four-point bending test.**

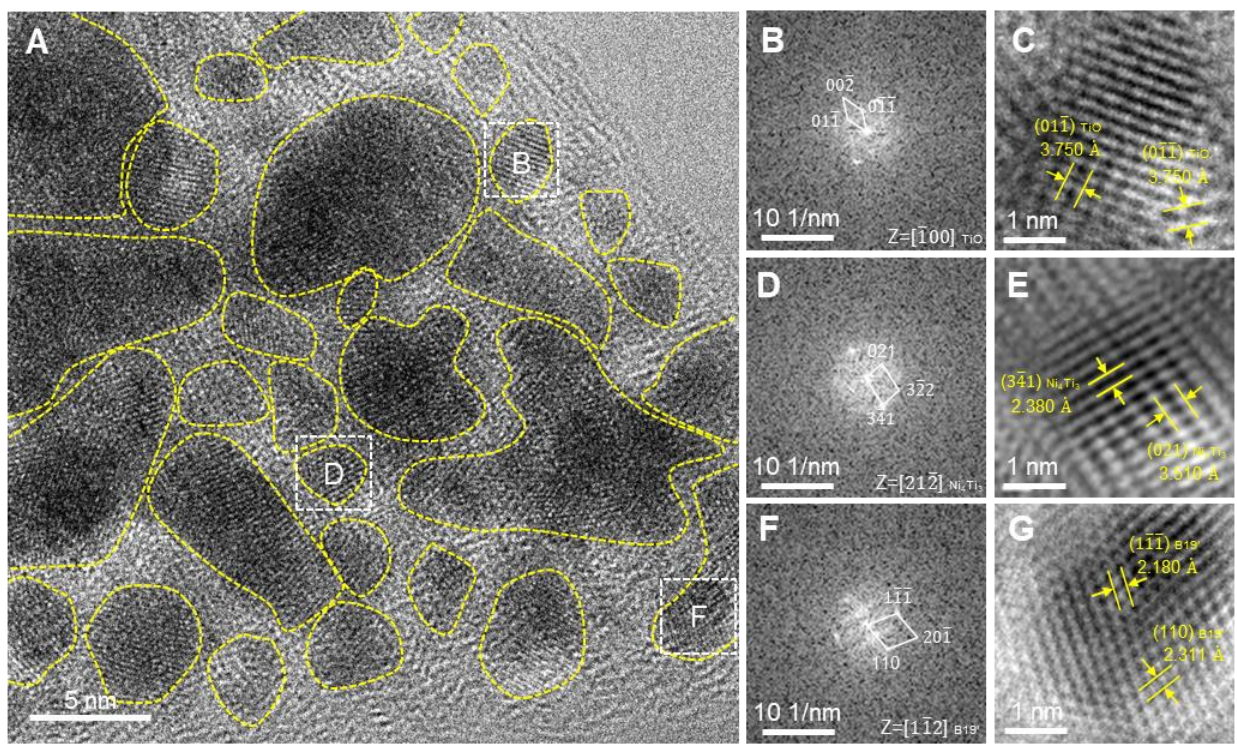

**Fig. S8.**
**High-resolution TEM characterization of the nanocomposite layer of the pw-LSP-treated surface**. (**A**) High-resolution TEM image of the pw-LSP-induced periodic surface structure. (**B**) Fast Fourier transform (FFT) pattern of the region B in (A) showing the presence of TiO. (**C**) Inverse FFT image of the TiO. (**D**) FFT pattern of the region D in (A) showing presence of $Ni_4Ti_3$. (**E**) Inverse FFT image of the $Ni_4Ti_3$. (**F**) FFT pattern of the region F in (A) showing the presence of B19′. (**G**) Inverse FFT image of the B19′.

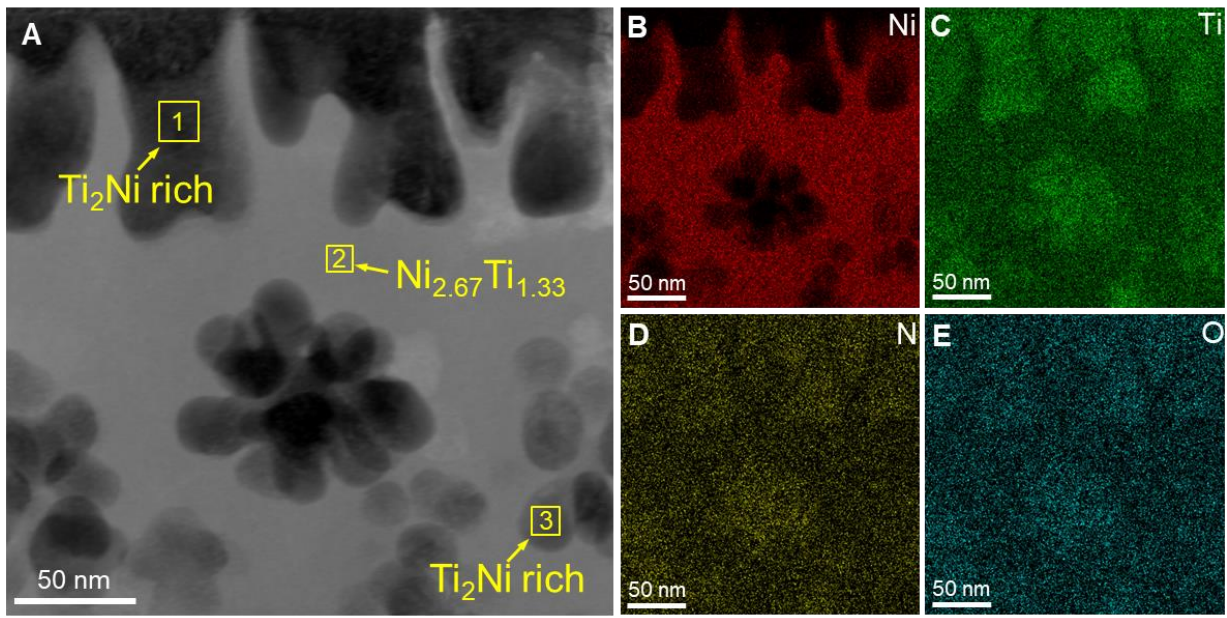

**Fig. S9.**
**Morphology and composition of the dendrite region of the pw-LSP treated surface.** (**A**) HAADF-STEM image. (**B-E**) The corresponding EDS elemental maps of (A). Quantitative EDS analysis (Table S1) reveals that the dendrites are $Ti_2Ni$ while the matrix is $Ni_{2.67}Ti_{1.33}$.

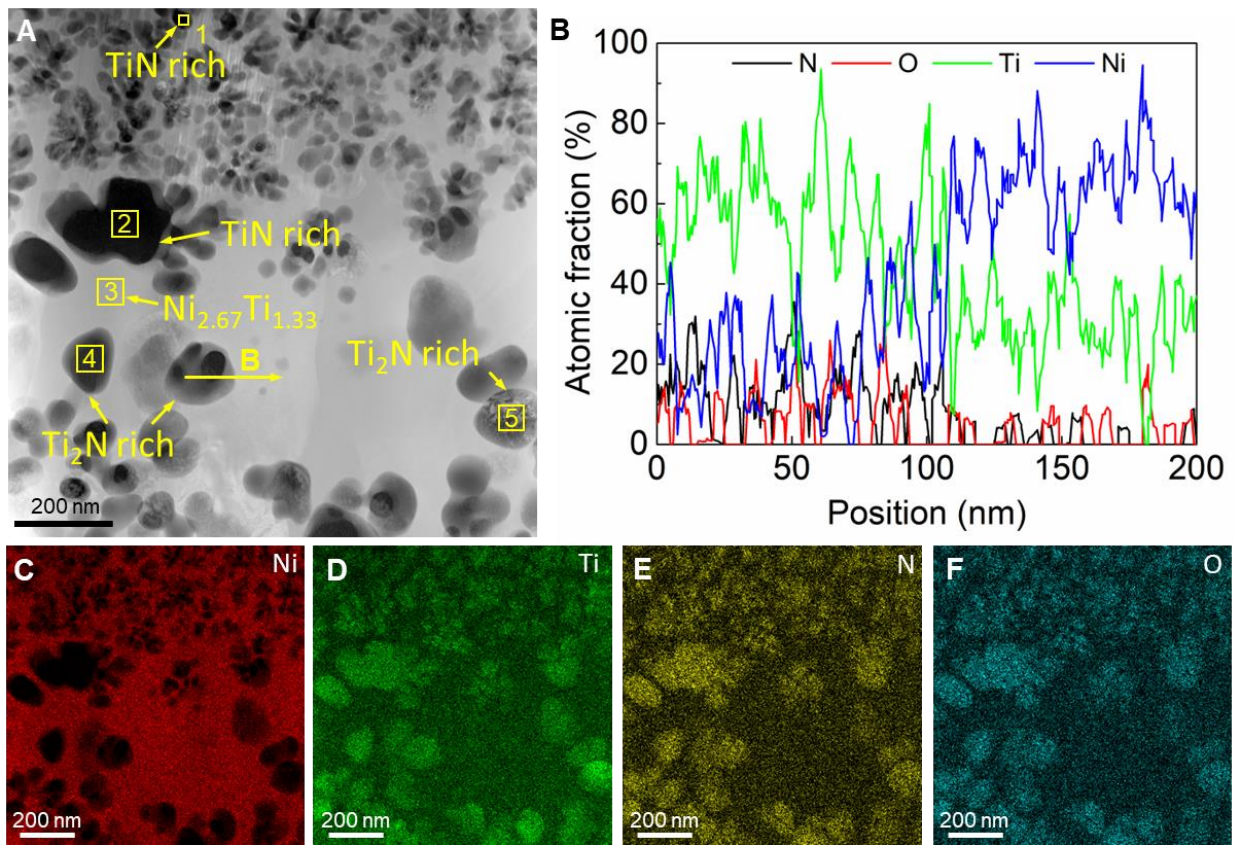

**Fig. S10.**

**Morphology and composition of the nitride layer.** (**A**) HAADF-STEM image. (**B**) EDS line profile across a precipitate/matrix boundary marked in (A). (**C-F**) The corresponding EDS elemental maps of (A). EDS elemental maps show that the nanoprecipitates are rich in Ti, O, and N, while the matrix is rich in Ni. Quantitative EDS analysis (Table S2) reveals that the snowflake-like precipitates are TiN, the spherical precipitates are $Ti_2N$, and the matrix is $Ni_{2.67}Ti_{1.33}$.

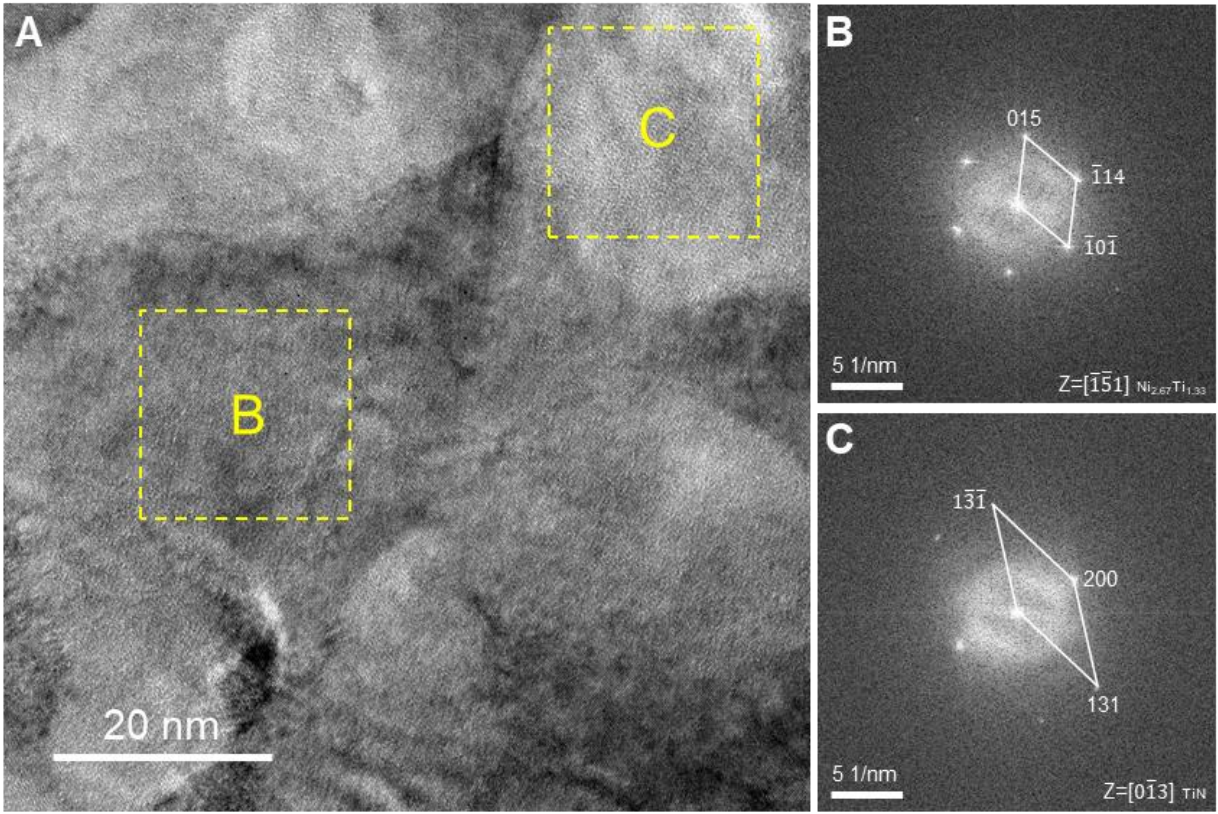

**Fig. S11.**
**HRTEM characterization of the TiN-rich region at the pw-LSP-treated sub-surface.** (**A**) HRTEM image of the pw-LSP-induced nitride layer. (**B**) FFT image of region B in (A). (**C**) FFT image of region C in (A). It is seen that there exist $Ni_{2.67}Ti_{1.33}$ phase in region B, and TiN phase in region C.

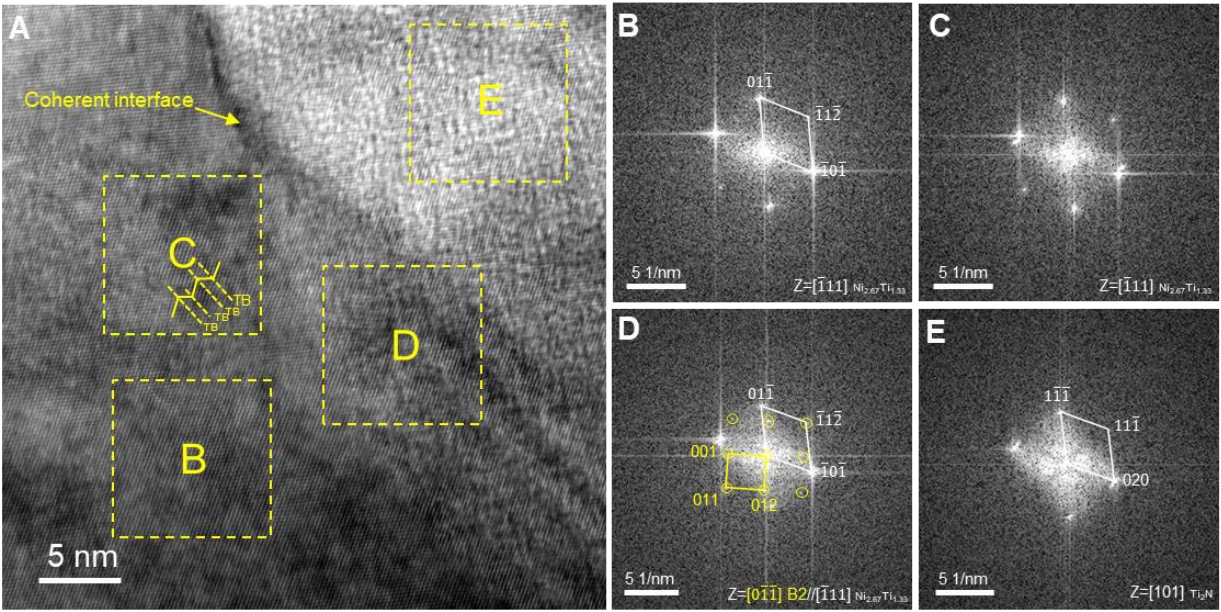

**Fig. S12.**

**HRTEM characterization of the Ti$_2$N-rich region at the pw-LSP-treated sub-surface.** (**A**) HRTEM image of the pw-LSP-induced nitride layer. (**B**) FFT image of region B in (A). (**C**) FFT image of region C in (A). (**D**) FFT image of region D in (A). (**E**) FFT image of region E in (A). It is seen that there exist Ni$_{2.67}$Ti$_{1.33}$ phase in region B, Ni$_{2.67}$Ti$_{1.33}$ phase can also have twinning in region C, Ni$_{2.67}$Ti$_{1.33}$ phase and B2 phase coexist in region D, and Ti$_2$N phase exists in region E.

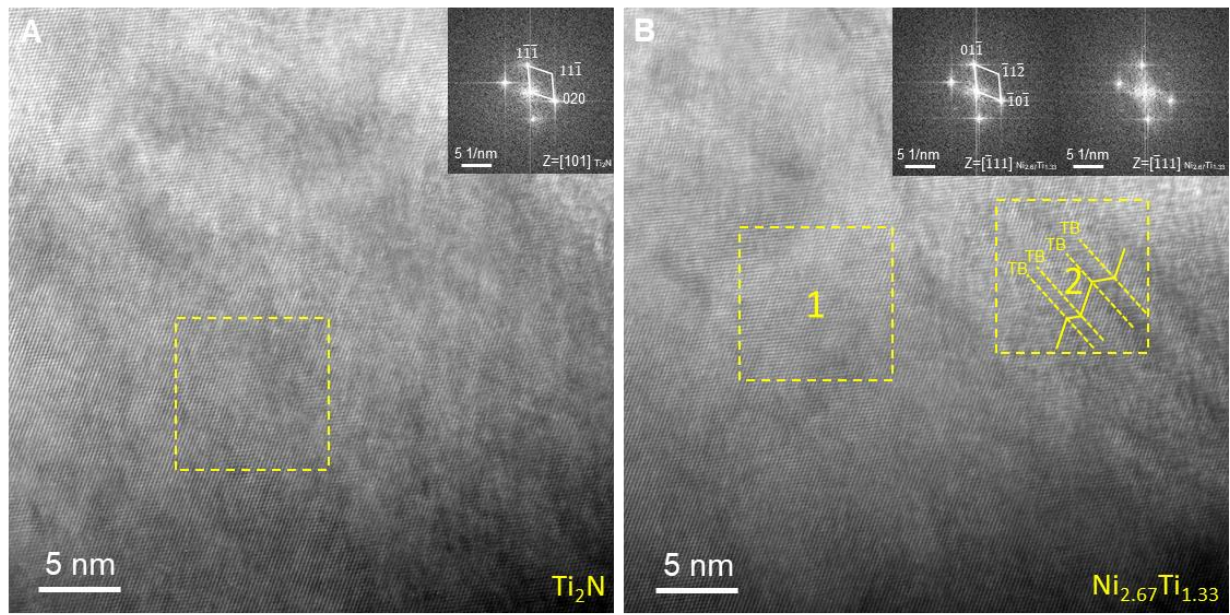

**Fig. S13.**
**HRTEM study of the Ti$_2$N and Ni$_{2.67}$Ti$_{1.33}$ phases.** (**A**) HRTEM image of Ti$_2$N phase, the inset is the corresponding FFT image of the selected region. (**B**) HRTEM image of Ni$_{2.67}$Ti$_{1.33}$ phase; the inset is the corresponding FFT image of the selected region.

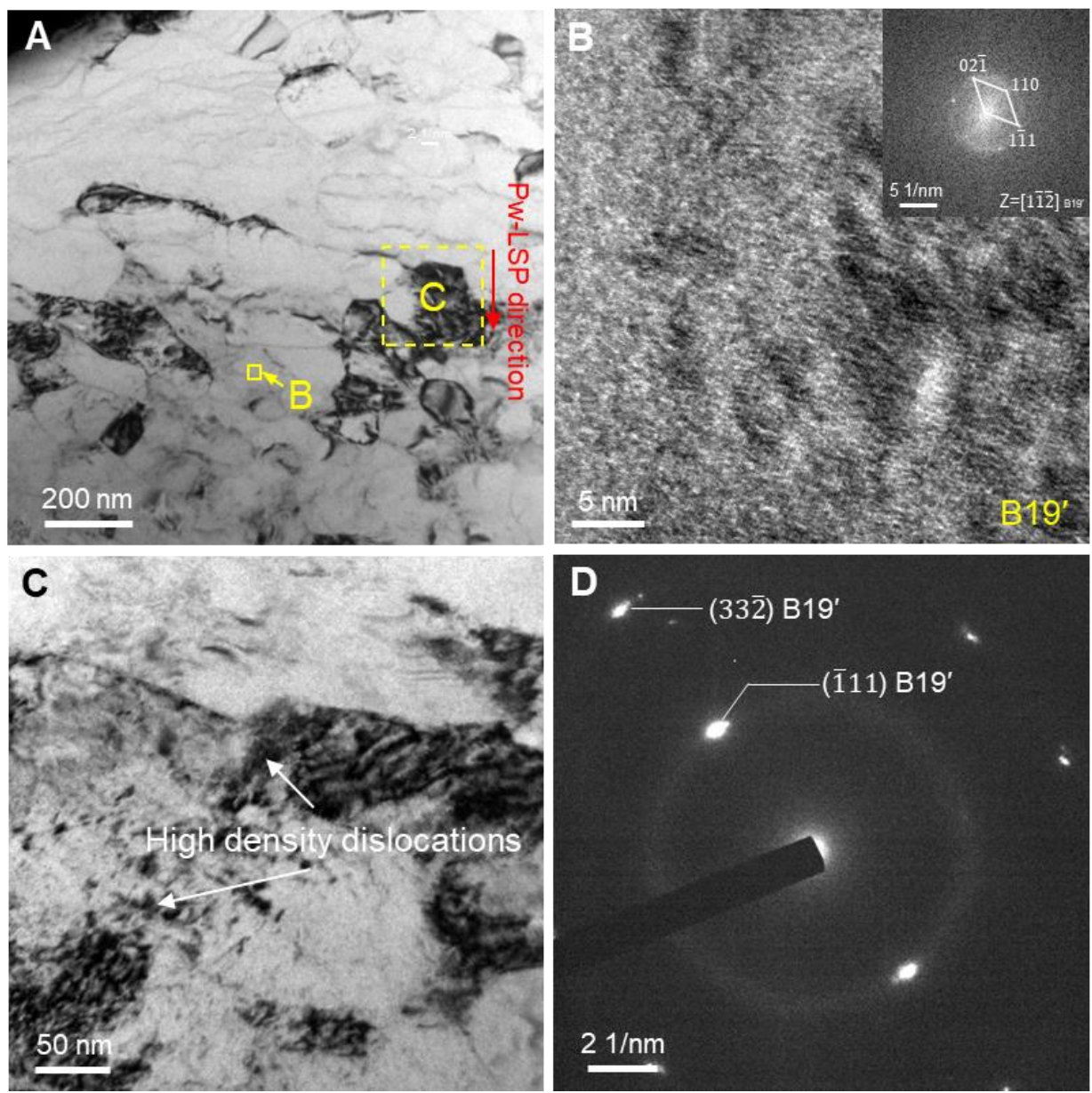

**Fig. S14.**

**TEM study of the region at a depth of 13 μm from pw-LSP treated top surface.** (**A**) Bright-field TEM image. The inset is SAED pattern of region a, indicating that region a is almost amorphous. (**B**) Enlarged TEM image of region B in (A). (**C**) SAED pattern of (B), indicating the existence of B19′ and R phases. (**D**) HRTEM of the interface region between amorphous and crystalline in A.

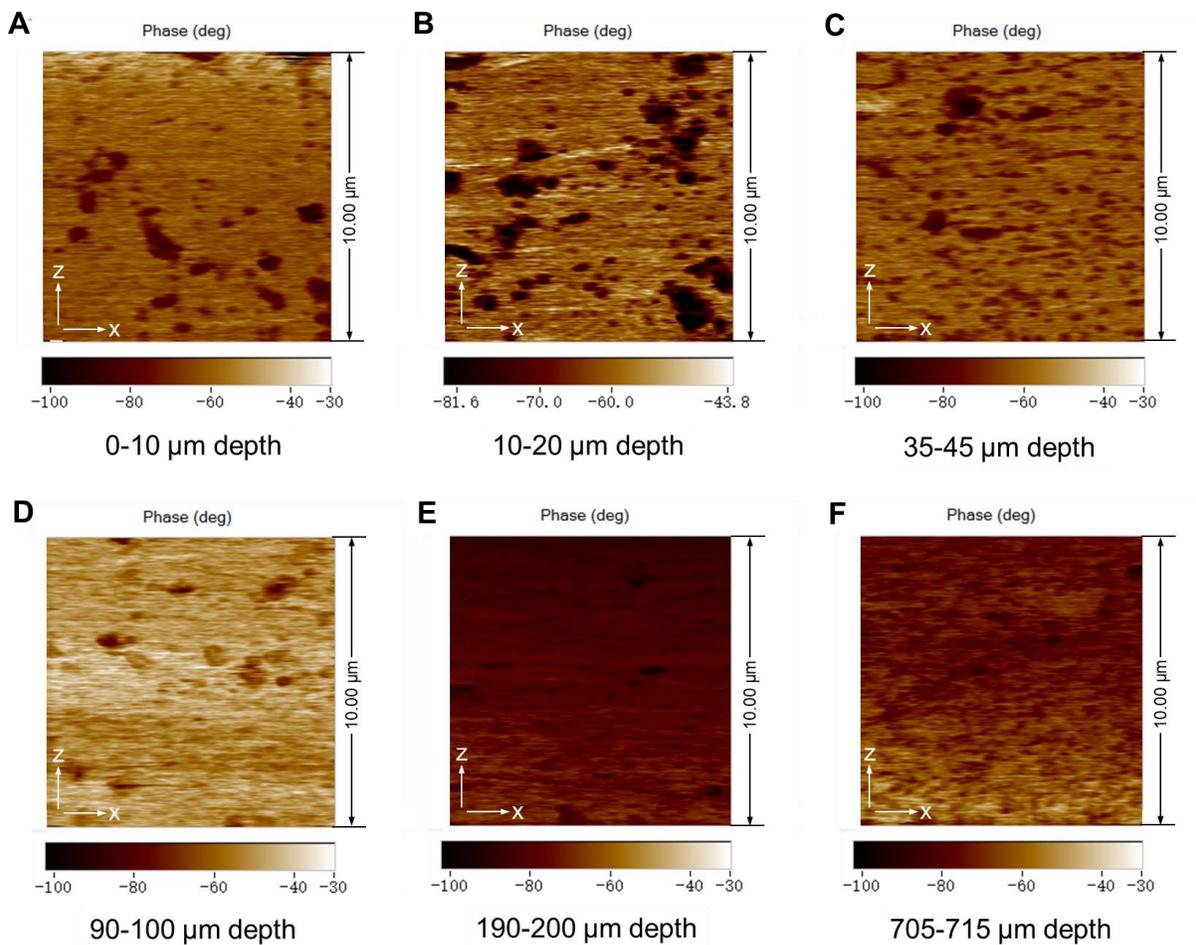

**Fig. S15.**
**Phase mapping of the cross-section of hierarchical NiTi at different depths from the pw-LSP treated top surface.** (**A**) 0-10 μm depth. (**B**) 10-20 μm depth. (**C**) 35-45 μm depth. (**D**) 90-100 μm depth. (**E**) 190-200 μm depth. (**F**) 705-715 μm depth.

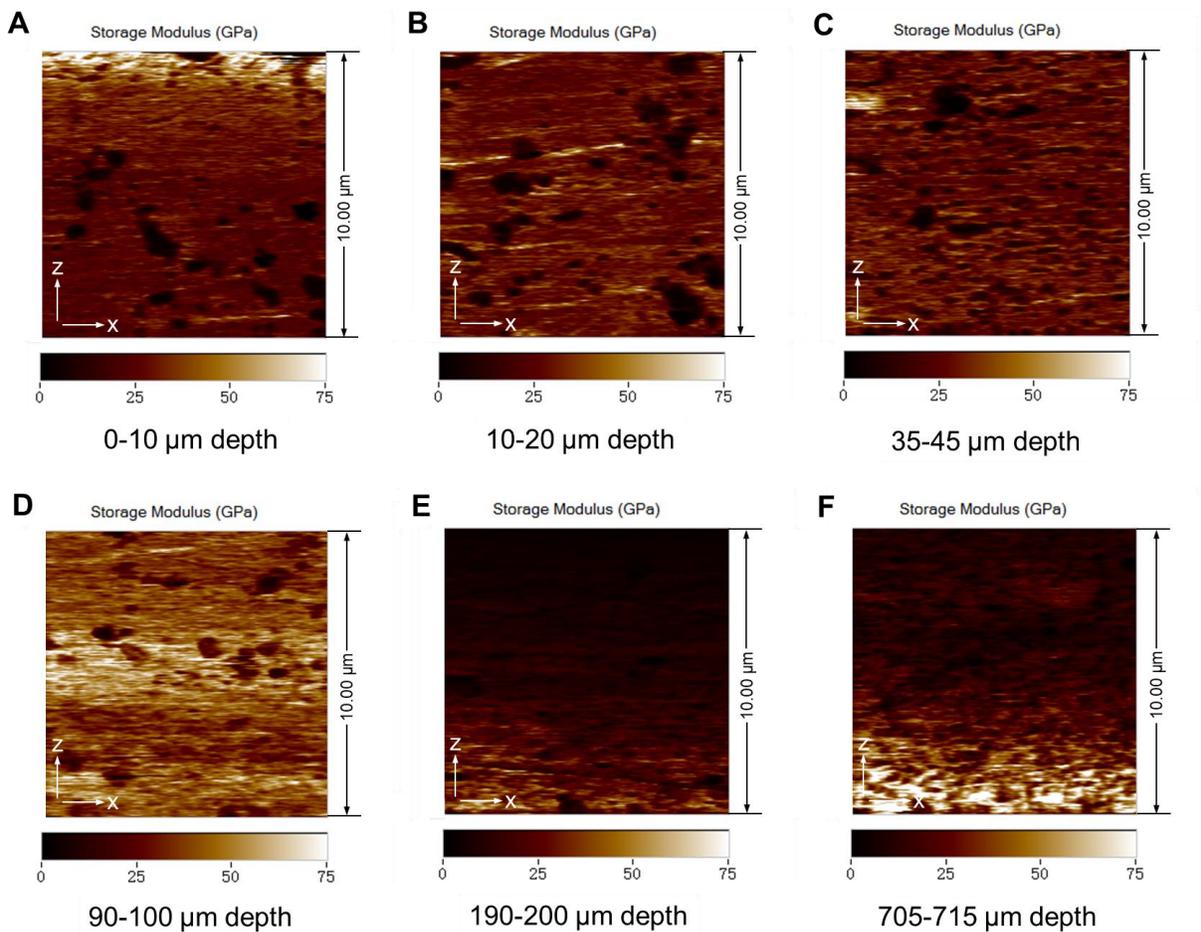

**Fig. S16.**
**Modulus mapping of the cross-section of hierarchical NiTi at different depths from the pw-LSP treated top surface.** (**A**) 0-10 μm depth. (**B**) 10-20 μm depth. (**C**) 35-45 μm depth. (**D**) 90-100 μm depth. (**E**) 190-200 μm depth. (**F**) 705-715 μm depth.

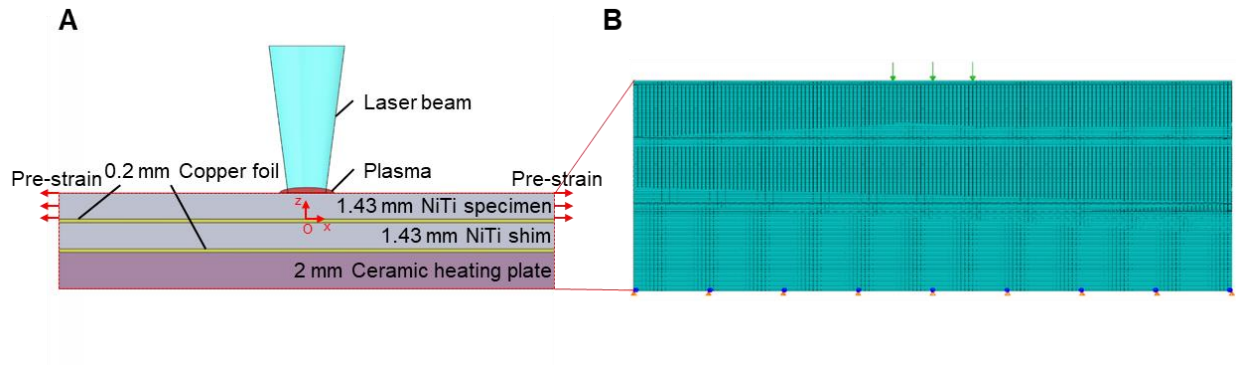

**Fig. S17.**
**2D Schematic representation of the pw-LSP treatment and FEM simulation model. (A)** Illustration of the 2D schematic for pw-LSP treatment. **(B)** Schematic representation of the FEM simulation mesh model and the corresponding boundary conditions.

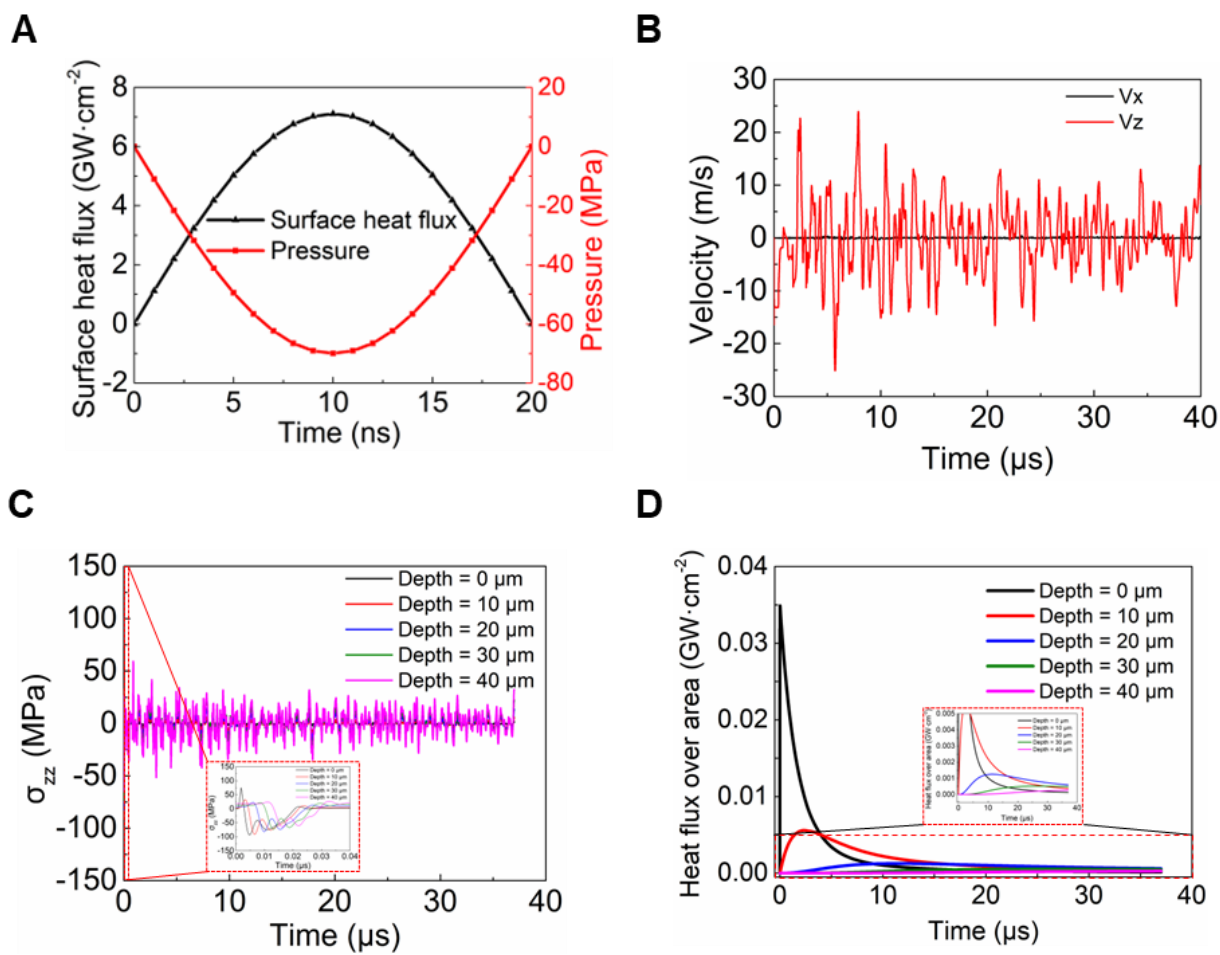

**Fig. S18.**
**FEM simulation results of pw-LSP treatment.** (**A**) The input surface heat flux-time and pressure-time curves in FEM simulation of pw-LSP treatment. (**B**) Calculated velocity-time curves of pw-LSP treated top surface. (**C**) Calculated $\sigma_{zz}$-time curves of different depths from the pw-LSP treated top surface. (**D**) Calculated heat flux-time curves of different depths from the pw-LSP treated top surface.

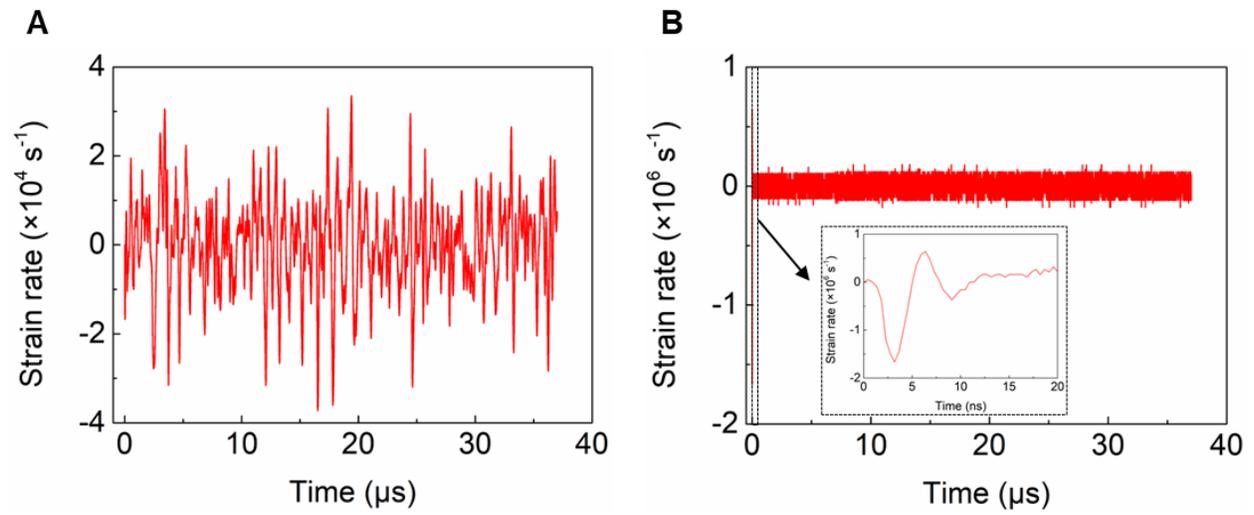

**Fig. S19.**
**Simulated strain rate-time curves at the pw-LSP treated top surface during pw-LSP treatment.** (**A**) x direction. (**B**) z direction.

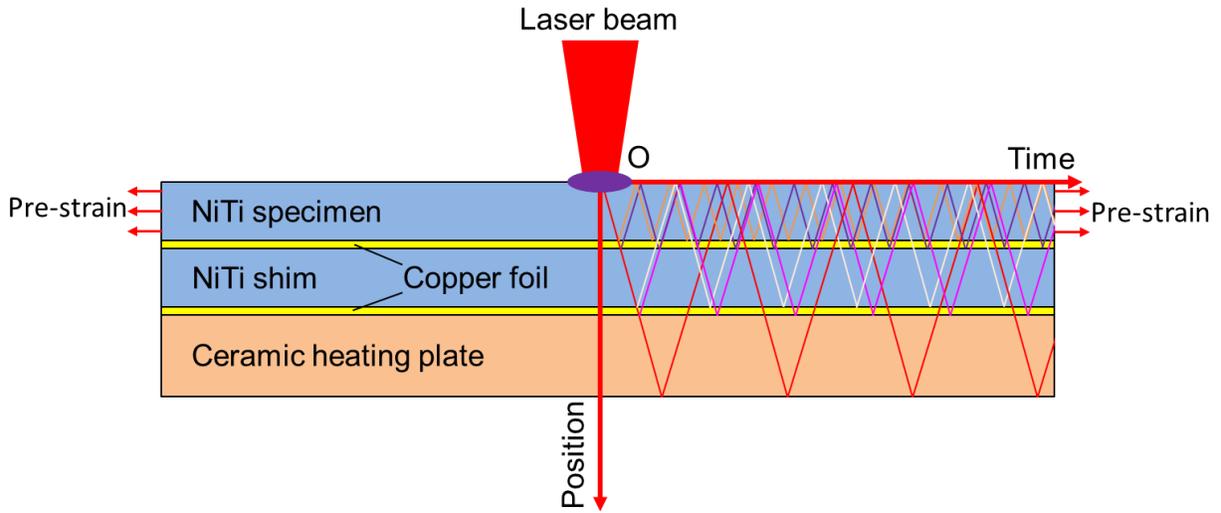

**Fig. S20.**

**Space-time diagram of the stress wave propagation in the specimen during the pw-LSP treatment.**

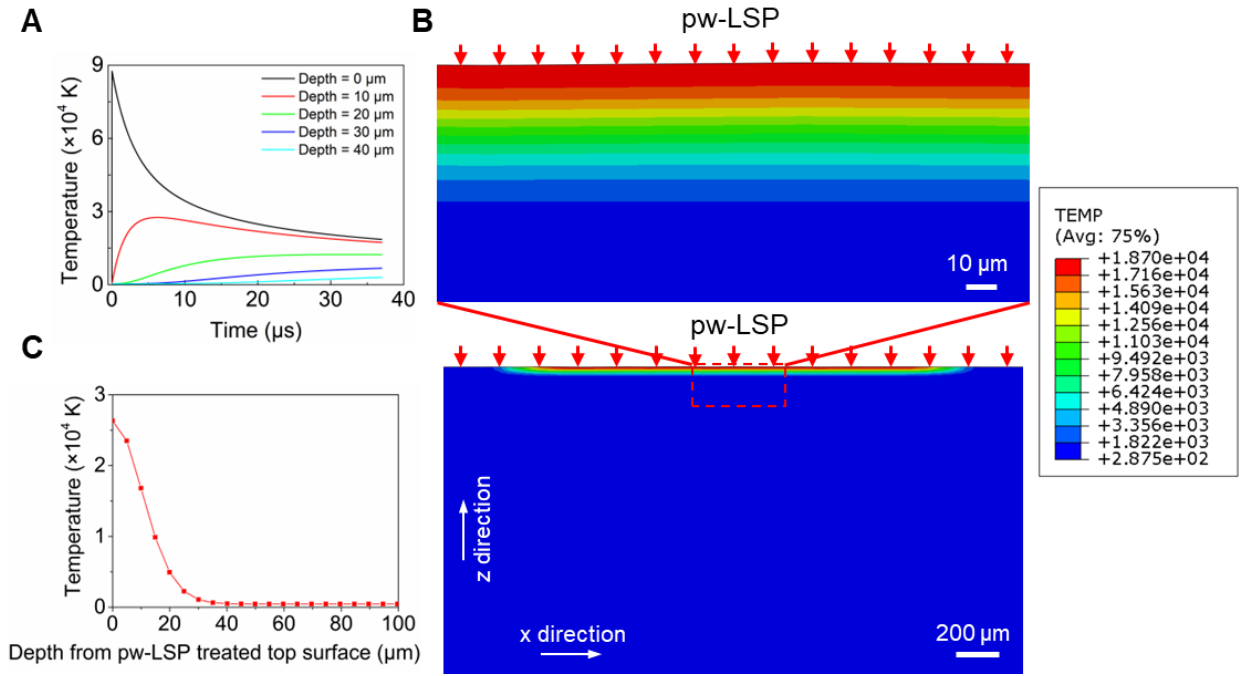

**Fig. S21.**
**Temperature calculation during pw-LSP treatment.** (**A**) Calculated temperature-time curves of NiTi surface region at different depths. (**B**) Calculated temperature field of NiTi surface after pw-LSP treatment. It is seen that the maximum surface temperature is about $1.3 \times 10^5$ K. (**C**) Temperature profile of NiTi specimen cross-section at $3.7 \times 10^{-5}$ s.

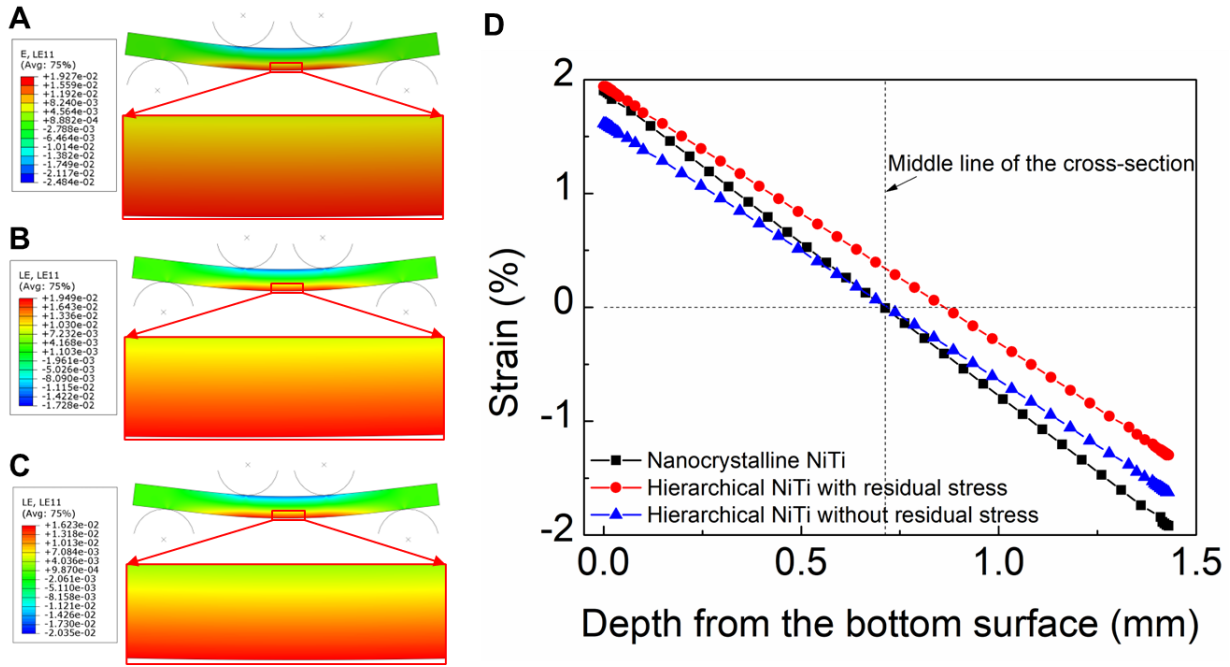

**Fig. S22.**
**Calculation of surface tensile strain along the *x*-direction by FEM.** (**A**) Calculated strain field of nanocrystalline NiTi along the *x*-direction, with a maximum surface tensile strain of about 1.927% at the bottom tensile surface, and a minimum surface compressive strain of about -2.484% at the top compressive surface. (**B**) Calculated strain field of hierarchical NiTi with residual stress along the *x*-direction, with a maximum surface tensile strain of about 1.943% at the bottom tensile surface, and a minimum surface compressive strain of about -1.728% at the top compressive surface. (**C**) Calculated strain field of hierarchical NiTi without residual stress along the *x*-direction, with a maximum surface tensile strain of about 1.623% and a minimum surface compressive strain of about -2.035%.

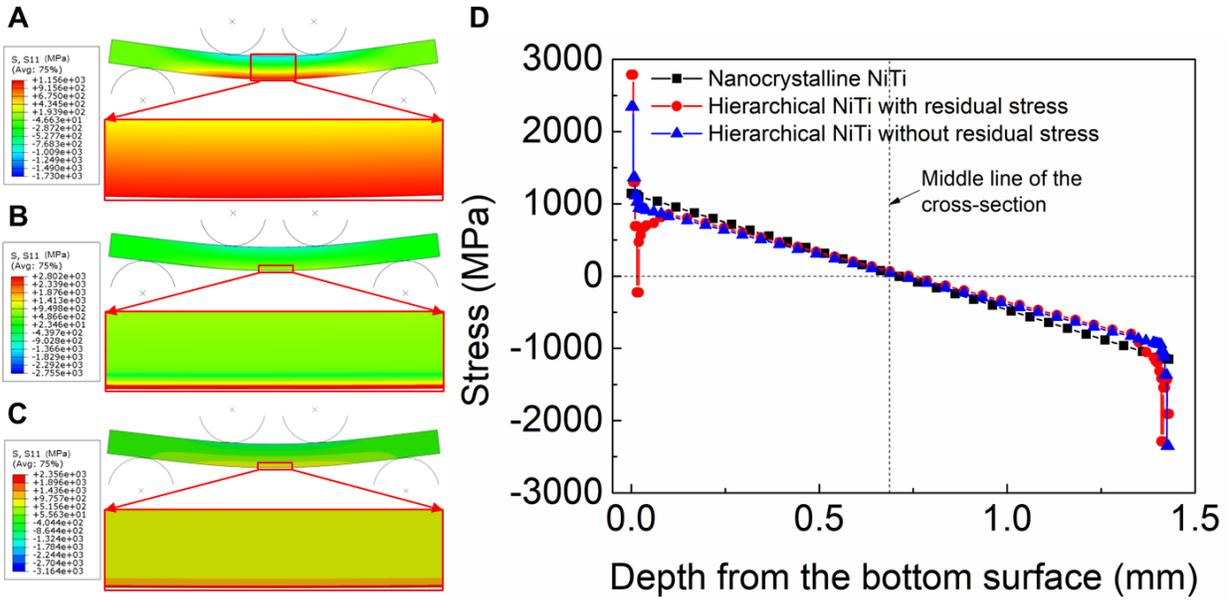

**Fig. S23.**

**Calculation of surface tensile stress along the *x*-direction by FEM.** (**A**) Calculated stress field of nanocrystalline NiTi along the *x*-direction, with a maximum surface tensile stress of about 1.156 GPa at the bottom tensile surface, and a minimum surface compressive stress of about -1.730 GPa at the top compressive surface. (**B**) Calculated strain field of hierarchical NiTi with residual stress along the *x*-direction, with a maximum surface tensile stress of about 2.802 GPa at the bottom tensile surface, and a minimum surface compressive stress of about -2.755 GPa at the top compressive surface. (**C**) Calculated strain field of hierarchical NiTi without residual stress along the *x*-direction, with a maximum surface tensile stress of about 2.356 GPa and a minimum surface compressive strain of about -3.164 GPa.

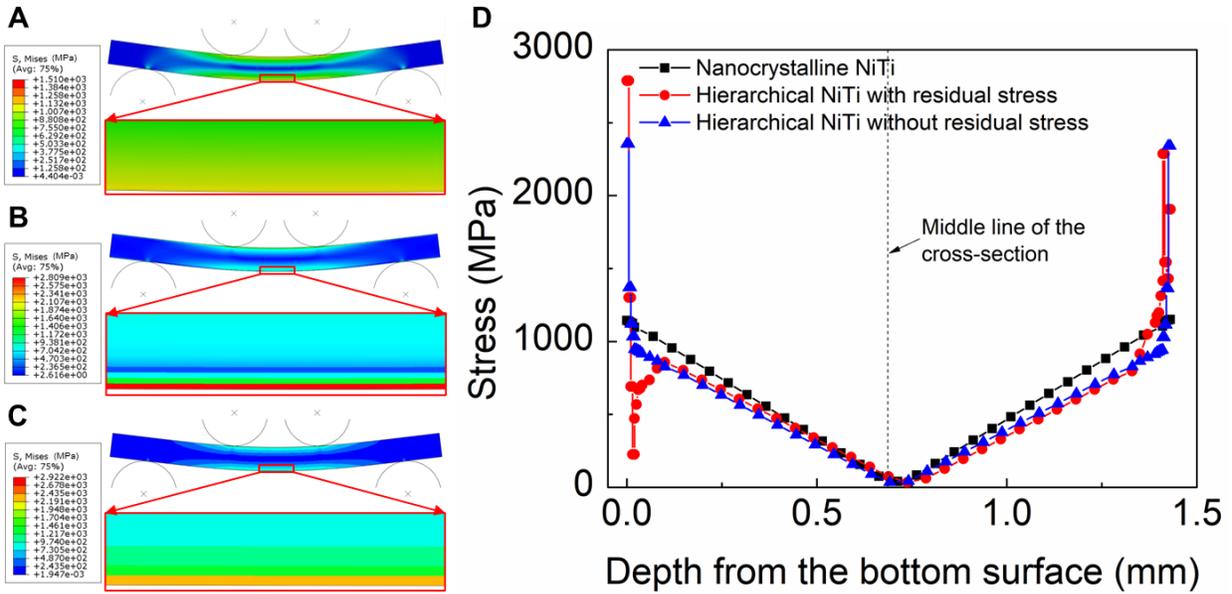

**Fig. S24.**
**Calculation of mises stress by FEM.** (**A**) Calculated sample mises stress field of nanocrystalline NiTi. (**B**) Calculated sample mises stress field of the hierarchical NiTi with residual stress. (**C**) Calculated sample mises stress field of the hierarchical NiTi without residual stress.

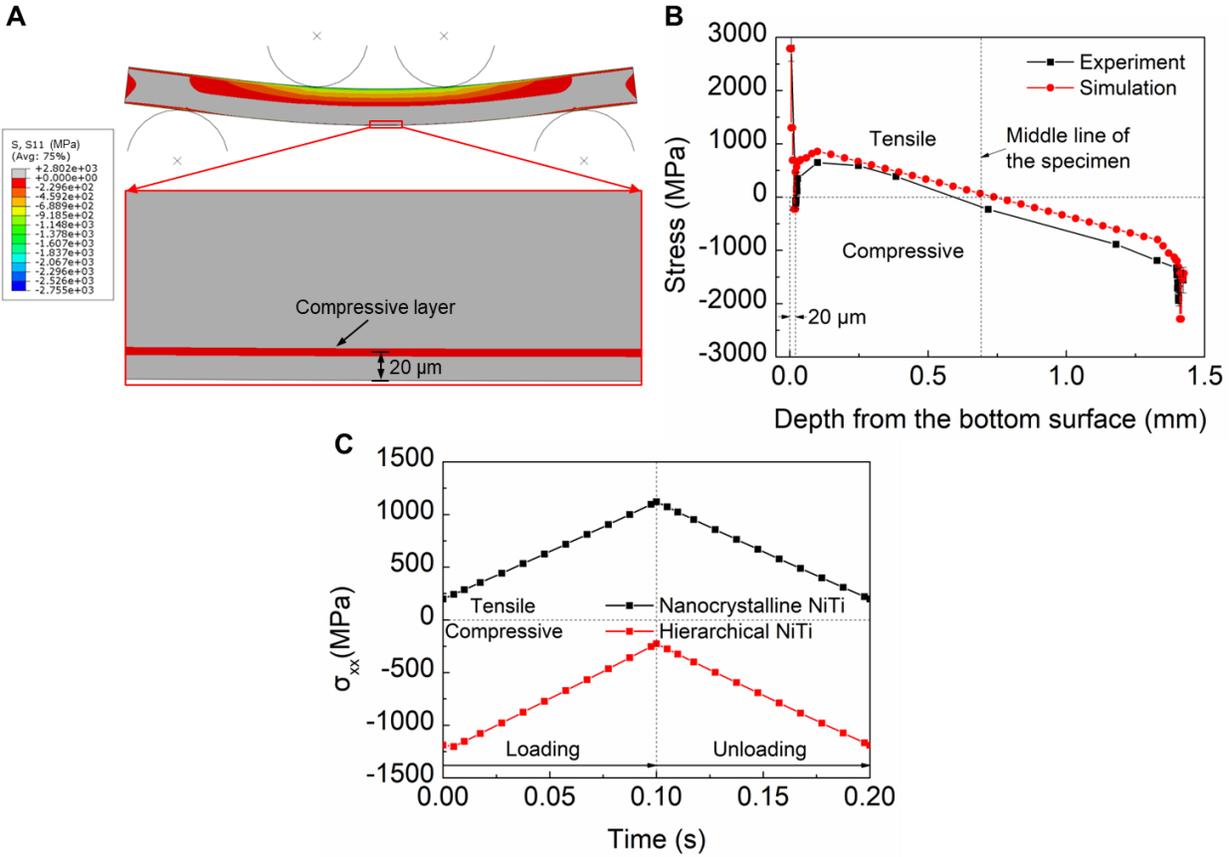

**Fig. S25.**
**Comparison of the stress field between experiment and simulation results.** (**A**) Calculated stress field of hierarchical NiTi along the *x*-direction, only compressive region is shown. (**B**) Comparison between experiment and simulation results of the stress field along the *x*-direction. (**C**) Comparison of the $\sigma_{xx}$ at ~20 μm depth from the tensile surfaces of nanocrystalline and hierarchical NiTi during loading and unloading.

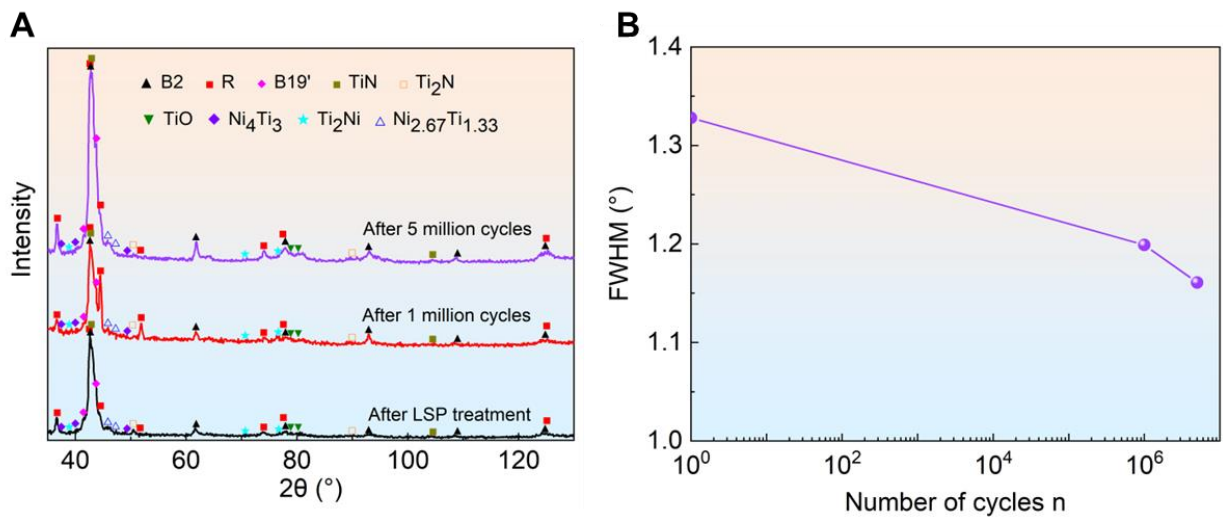

**Fig. S26.**
(**A**) XRD patterns of the hierarchical NiTi surface (tensile side) before and after cyclic bending. (**B**) Change of FWHM with cycle number. It is observed that the FWHM decreases with an increase in the cycle number.

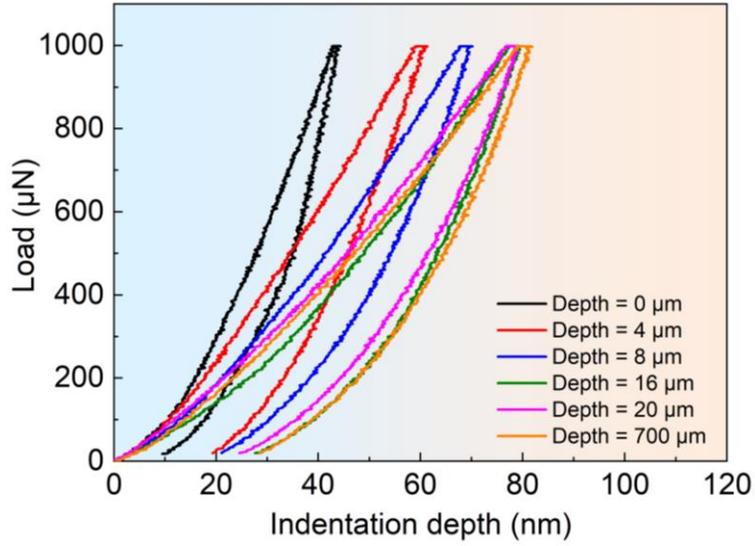

**Fig. S27.**

**Load-indentation depth curves of the cross-section of hierarchical NiTi.** It reveals a consistent pattern of increasing indentation depth from the top surface to a depth of 20 μm, which remains stable with further depth increments.

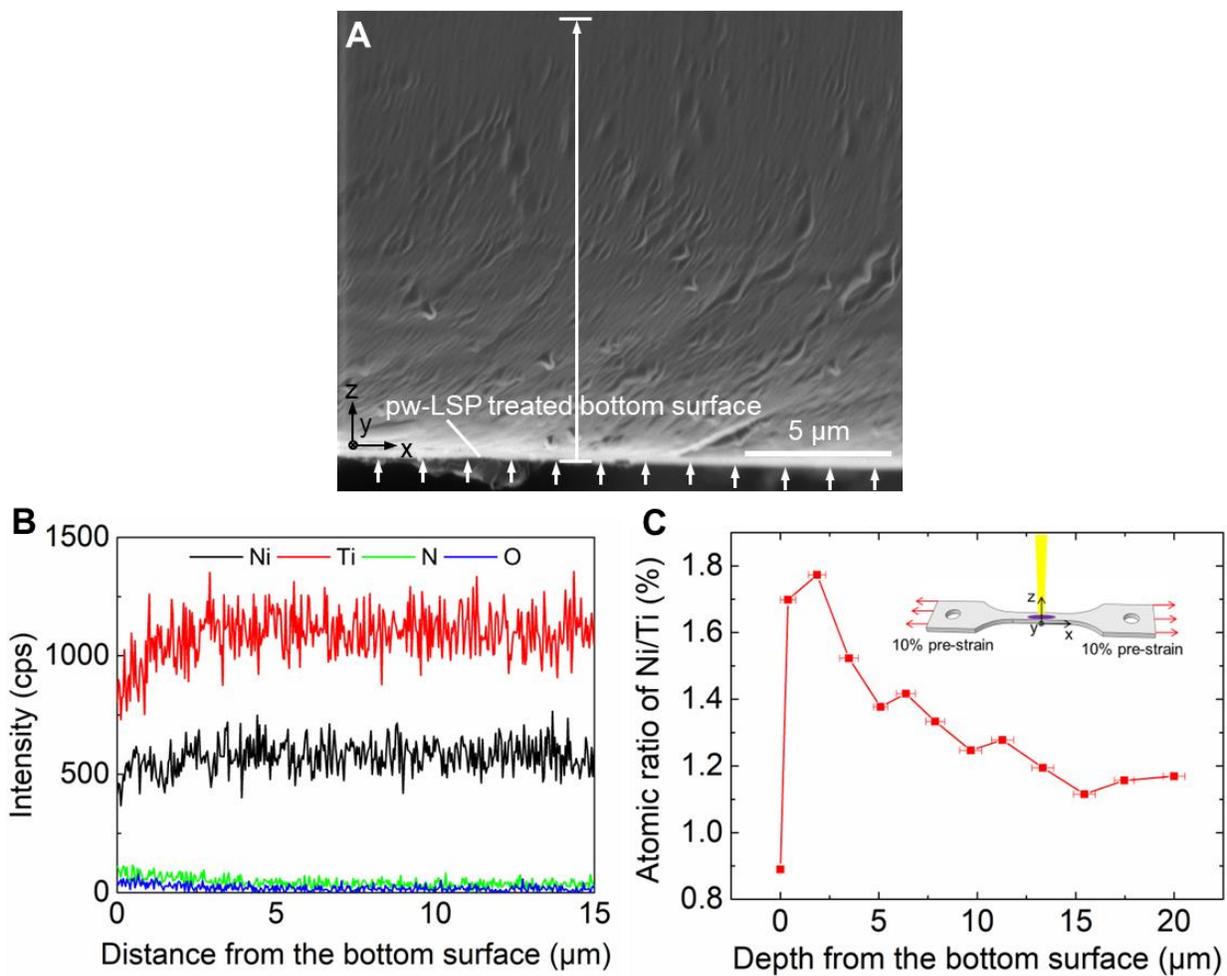

**Fig. S28.**
**SEM image (A), EDS line profiles along the depth (B), and Ni/Ti atomic ratio of the pw-LSP-treated specimen cross-section.**

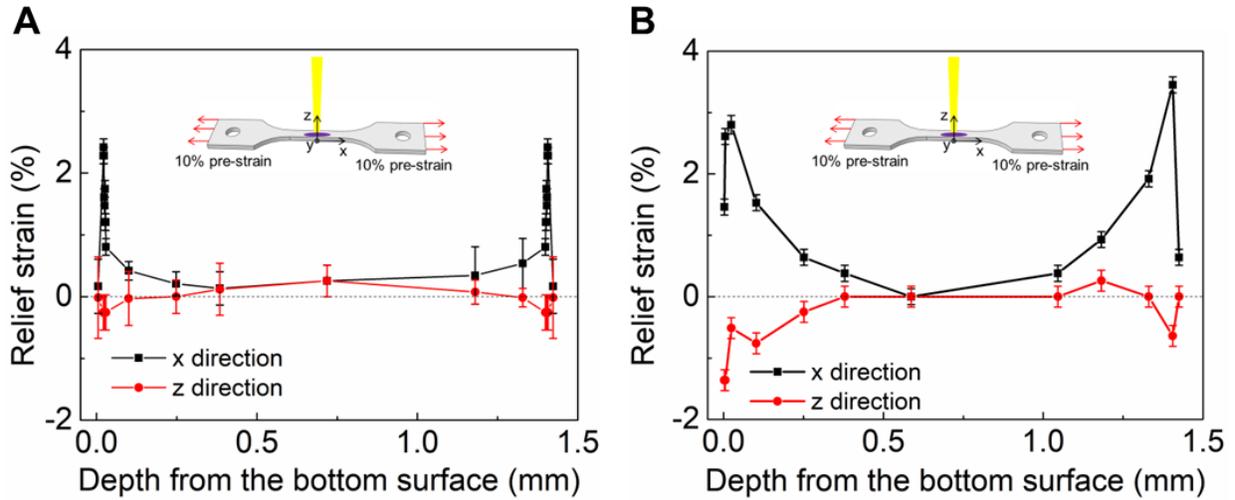

**Fig. S29.**
**Distribution of relief strain before and after bending fatigue test of 5 million cycles at different depths of the LSP-treated surface.** (**A**) Relief strain before bending fatigue test. (**B**) Relief strain after bending fatigue test. The relief strain before fatigue test shows a gradient distribution along the *x*-direction, reaching a maximum of about 2.4% at approximately 20 μm depth from the pw-LSP treated bottom surface. Along the *z*-direction, the relief strain is negative at around 100 μm depth from the pw-LSP treated bottom surface and becomes positive in the inner region of the pw-LSP treated sample. The relief strain after fatigue test also shows a gradient distribution along the *x*-direction, reaching a maximum of about 2.8% at approximately 25 μm depth from the pw-LSP treated bottom surface. Along the *z*-direction, the relief strain is negative in almost the entire depth of the NiTi specimen.

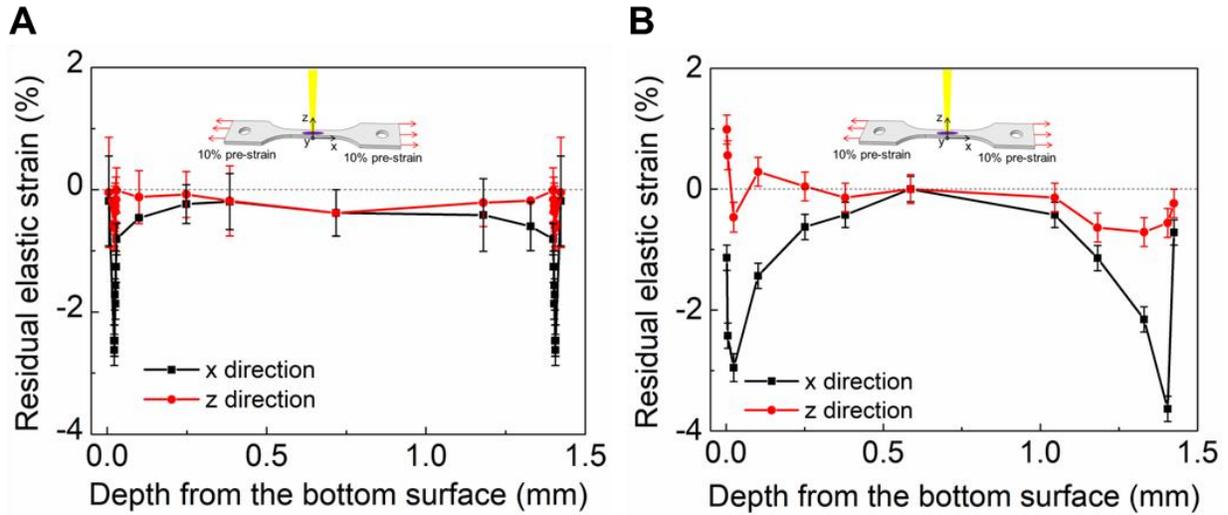

**Fig. S30.**
**Distribution of residual elastic strain of specimen before and after bending test of 5 million cycles at different depths of the LSP-treated surface.** (**A**) Residual elastic strain before bending fatigue test. (**B**) Residual elastic strain after bending fatigue test. The pre-fatigue residual elastic strain exhibits a gradient distribution along the x-direction, with a minimum value of approximately -2.7% observed at a depth of around 20 μm from the bottom surface treated with pw-LSP. Additionally, along the z-direction, the residual elastic strain remains negative across nearly all depths of the pw-LSP treated sample. Following the fatigue test, the residual elastic strain displays a gradient distribution along the x-direction. At the tensile side, the strain reaches a minimum value of about -3.0% at approximately 24 μm depth from the bottom surface treated with pw-LSP and a minimum value of about -3.6% at approximately 25 μm depth from the top surface treated with pw-LSP. Along the z-direction, the residual elastic strain measures about 1.0% at the bottom surface treated with pw-LSP and transitions to negative values within the inner region of the treated sample.

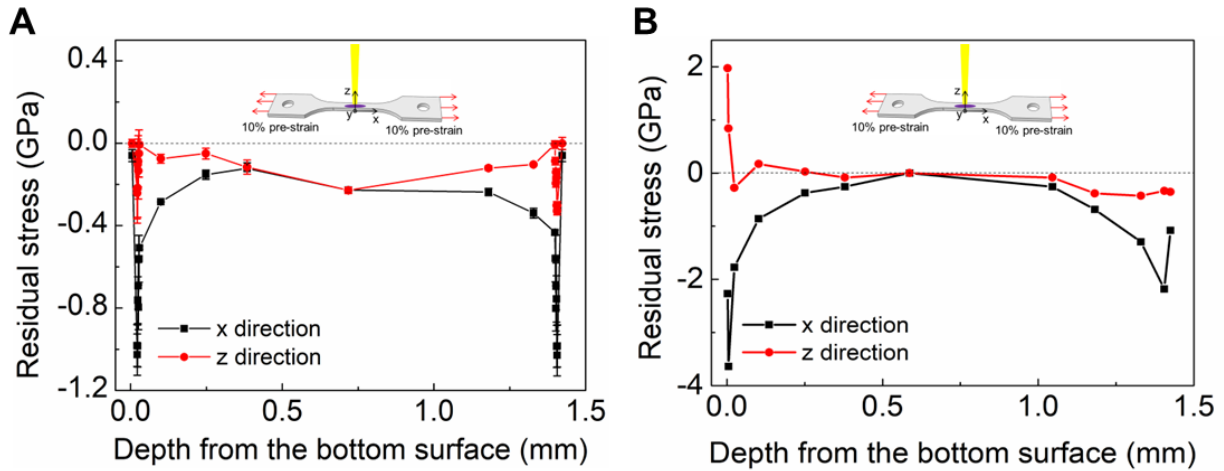

**Fig. S31.**
**Distribution of residual stress of specimen before and after bending test of 5 million cycles at different depths of the LSP-treated surface.** (**A**) Residual stress before bending fatigue test. (**B**) Residual stress after bending fatigue test. The pre-fatigue residual stress exhibits a gradient distribution along the x-direction, with a minimum value of approximately -1.0 GPa observed at a depth of around 20 μm from the bottom surface treated with pw-LSP. Additionally, along the z-direction, the residual elastic strain remains negative across nearly all depths of the pw-LSP treated sample. Following the fatigue test, the residual stress displays a gradient distribution along the x-direction. At the tensile side, the residual stress reaches a minimum value of about -3.6 GPa at approximately 5 μm depth from the bottom surface treated with pw-LSP and a minimum value of about -2.2 GPa at approximately 20 μm depth from the top surface treated with pw-LSP. Along the z-direction, the residual stress reaches about 2.0 GPa at the bottom surface treated with pw-LSP and transitions to negative values within the inner region of the treated sample.

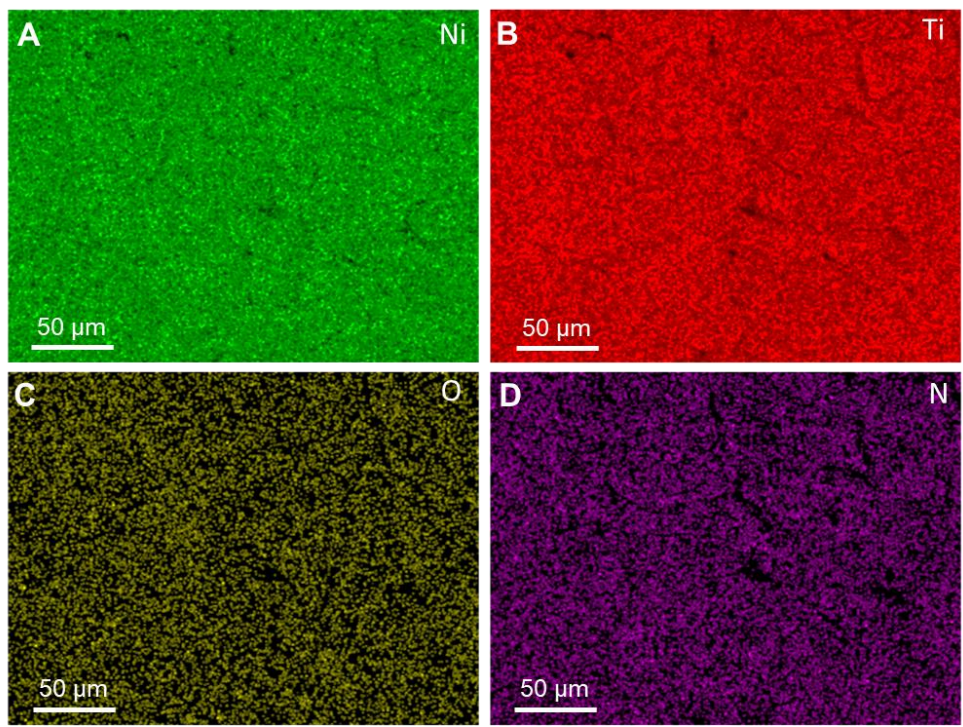

**Fig. S32.**
**EDS elemental mapping of the pw-LSP-treated surface.** (A) EDS elemental mapping of Ni element; (A) EDS elemental mapping of Ti element; (A) EDS elemental mapping of O element; (A) EDS elemental mapping of N element.

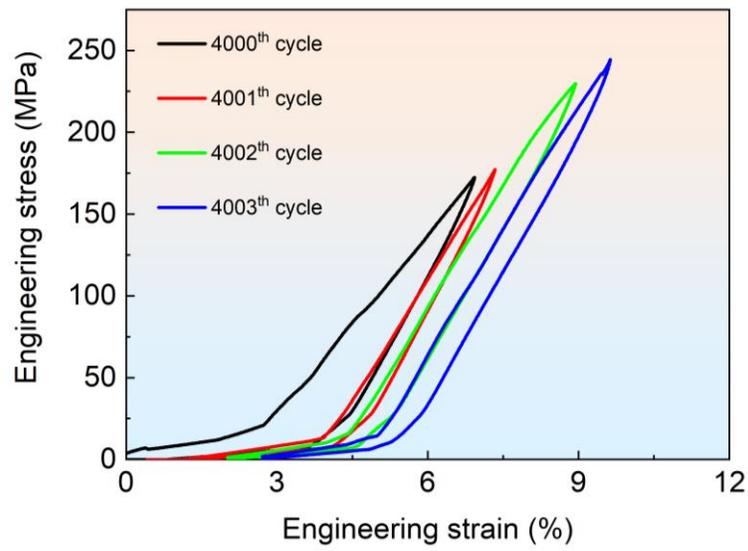

**Fig. S33.**

**Engineering stress-strain curves of 4000$^{th}$-4003$^{th}$ cycles under the displacement-controlled mode.** It is seen that the micro-tension sample can maintain superelastic behaviour under the maximum tensile engineering strain of 9%.

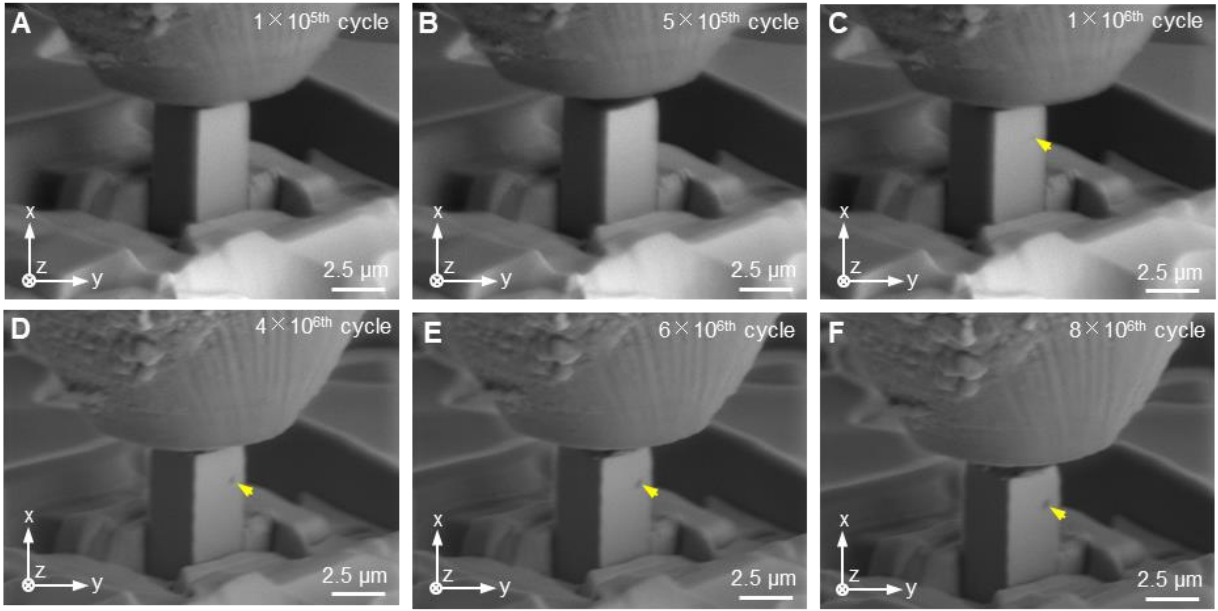

**Fig. S34.**
*In situ* **SEM images of the surface morphology of the cyclic-compression specimen at a depth of 20 μm.** (**A**) $1\times10^{5}$th cycle; (**B**) $5\times10^{5}$th cycle; (**C**) $1\times10^{6}$th cycle; (**D**) $4\times10^{6}$th cycle; (**E**) $6\times10^{6}$th cycle; (**F**) $8\times10^{6}$th cycle.

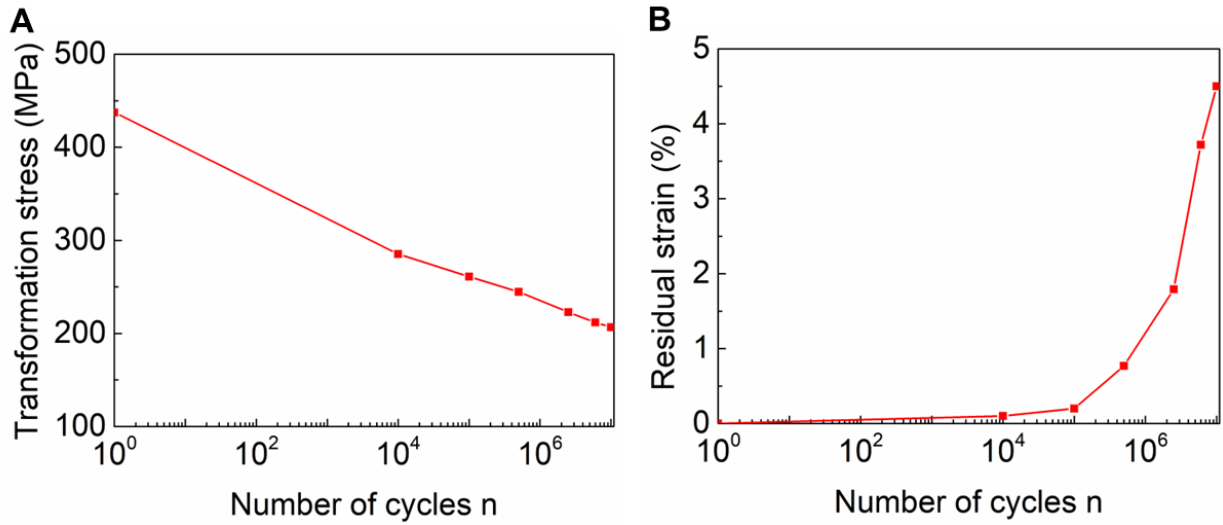

**Fig. S35.**
**Evolution of transformation stress and residual strain of the micron pillar after different number of cycles.** (**A**) Transformation stress; (**B**) Residual strain.

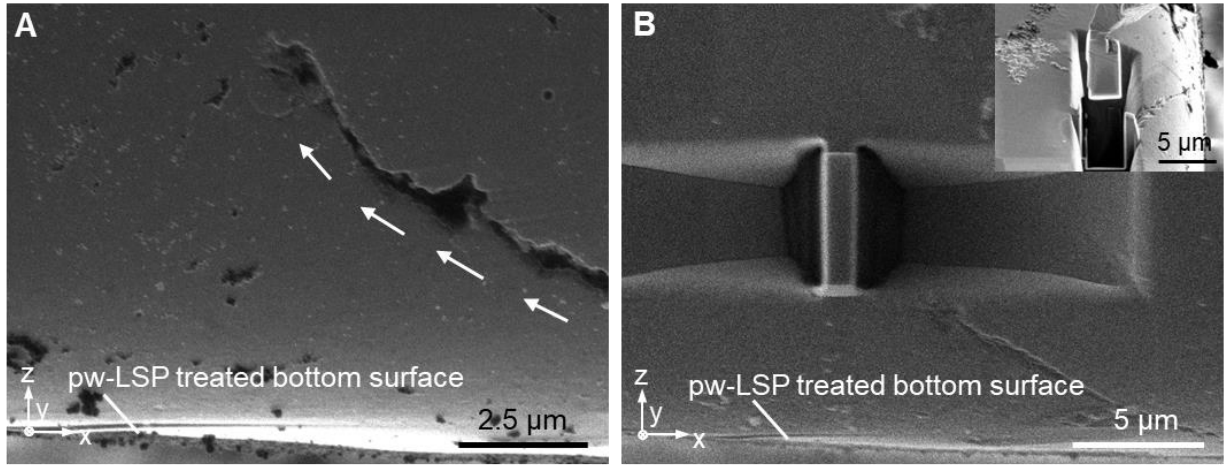

**Fig. S36.**

**Preparation of TEM sample at the fatigue crack tip with the lift-out method.** (**A**) SEM image of the fatigue crack tip at the bottom tensile surface after bending fatigue test of 5 million cycles; (**B**) SEM image of the TEM specimen at the crack tip, inset is the lift-out process.

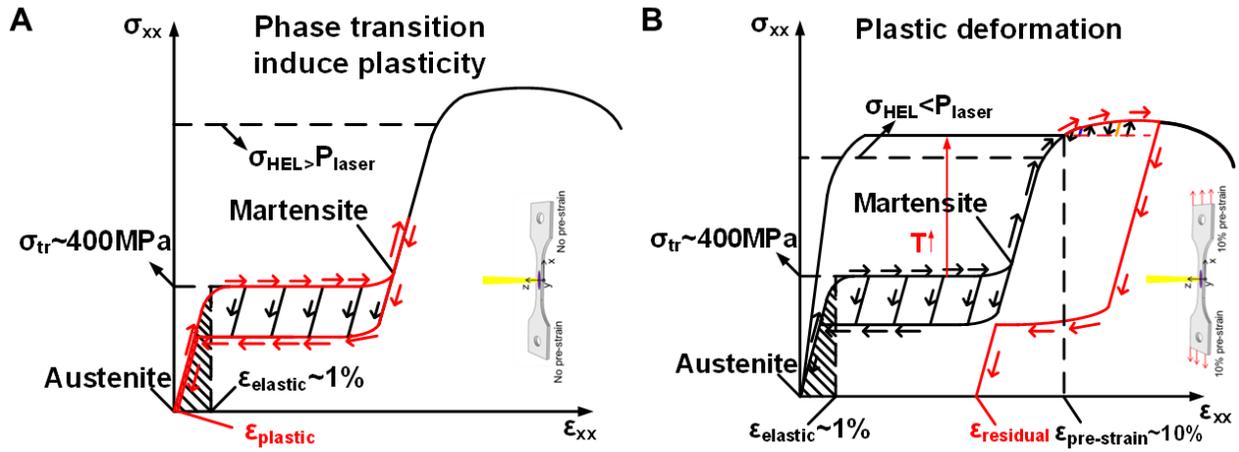

**Fig. S37.**
**Schematic of the deformation mechanism of NiTi treated under different conditions.** (**A**) Deformation mechanism of NiTi treated with LSP with no pre-strain and heating. (**B**) Deformation mechanism of NiTi treated with pw-LSP.

**Table S1.**

Quantitative analysis of atomic percentage of different regions at the pw-LSP treated top surface.

| Region | Ni (%) | Ti (%) | N (%) | O (%) |
|---|---|---|---|---|
| 1 | 19.90 ±3.17 | 61.20 ±4.68 | 7.91 ±1.54 | 11.00 ±2.31 |
| 2 | 63.90 ±4.80 | 29.60 ±4.52 | 0.05 ±0.05 | 6.45 ±1.42 |
| 3 | 33.00 ±5.92 | 59.50 ±5.01 | 0.04 ±0.04 | 7.46 ±2.29 |

**Table S2.**

Quantitative analysis of atomic percentage of different regions of the nitride layer.

| Region | Ni (%) | Ti (%) | N (%) | O (%) |
|---|---|---|---|---|
| 1 | 0.30 ±0.30 | 36.70 ±7.56 | 41.30 ±9.94 | 21.70 ±6.50 |
| 2 | 0.42 ±0.11 | 48.10 ±4.39 | 42.20 ±3.65 | 9.28 ±1.89 |
| 3 | 63.30 ±4.85 | 30.50 ±4.60 | 5.57 ±1.31 | 0.63 ±0.63 |
| 4 | 0.30 ±0.30 | 59.80 ±12.60 | 28.00 ±2.02 | 11.90 ±6.19 |
| 5 | 5.61 ±0.99 | 61.60 ±4.24 | 24.60 ±2.77 | 8.19 ±1.75 |

**Table S3.**

Related materials properties in the FEM simulation of pw-LSP treatment.

| Material | Density (kg/m$^3$) | Elastic modulus (GPa) | Poisson's ratio | Inelastic heat fraction | Thermal conductivity (W · m$^{-1}$ · K$^{-1}$) | Specific heat capacity (J · g$^{-1}$ · ℃$^{-1}$) |
|---|---|---|---|---|---|---|
| **NiTi** | 6450 | 60 | 0.30 | 0.9 | 10 | 0.320 |
| **Copper** | 7764 | 110 | 0.34 | 0.9 | 385 | 0.385 |
| **Al$_2$O$_3$** | 3960 | 370 | 0.22 | 0.9 | 32 | 0.920 |

**Table S4.**

Related parameters of Johnson-Cook model in the FEM simulation of pw-LSP treatment.

| Material | $A$ (MPa) | $B$ (MPa) | $N$ | $C$ | $M$ | $T_r$ (K) | $T_m$ (K) | $\dot{\varepsilon}_0$ (s$^{-1}$) |
|---|---|---|---|---|---|---|---|---|
| **NiTi** | 510 | 0 | 0 | 0.087 | 0.36 | 300 | 1583 | $5.6 \times 10^4$ |
| **Copper** | 90 | 292 | 0.31 | 0.025 | 1.09 | 300 | 1356 | 1 |